\documentclass[sigconf]{acmart}

\usepackage{algorithm}
\usepackage{array}
\usepackage[noend]{algpseudocode}
\usepackage{algpseudocode,algorithm}
\usepackage{balance}
\usepackage{multirow}
\usepackage{paralist}
\usepackage{subcaption}
\usepackage{colortbl}
\usepackage{hyperref}

\newtheorem{theorem}{Theorem}[section]

\newtheorem{definition}[theorem]{Definition}
\newtheorem{lemma}{Lemma}[theorem]

\AtBeginDocument{%
  \providecommand\BibTeX{{%
    Bib\TeX}}}

\setcopyright{acmlicensed}
\copyrightyear{2018}
\acmYear{2018}
\acmDOI{XXXXXXX.XXXXXXX}
\acmConference[Conference acronym 'XX]{Make sure to enter the correct
  conference title from your rights confirmation email}{June 03--05,
  2018}{Woodstock, NY}
\acmISBN{978-1-4503-XXXX-X/2018/06}

\begin{document}

\title{Triadic First-Order Logic Queries in Temporal Networks}

\author{Omkar Bhalerao}

\affiliation{%
  \institution{University of California, Santa Cruz}
  \city{Santa Cruz}
  \country{USA}
}
\email{obhalera@ucsc.edu}

\author{Yunjie Pan}
\affiliation{%
  \institution{University of Michigan, Ann Arbor}
  \city{Ann Arbor}
  \country{USA}}
\email{panyj@umic.edu}

\author{C. Seshadhri}
\affiliation{%
  \institution{University of California, Santa Cruz}
  \city{Santa Cruz}
  \country{USA}
}
\email{sesh@ucsc.edu}

\author{Nishil Talati}
\affiliation{%
 \institution{University of Michigan, Ann Arbor}
 \city{Ann Arbor}
 \country{USA}}
 \email{talatin@umich.edu}


\begin{abstract}
  Motif counting is a fundamental problem in network analysis, and there is a rich literature of theoretical and applied algorithms for this problem. Given a large input network $G$, a motif $H$ is a small 
    "pattern" graph indicative of special local structure. Motif/pattern mining involves finding all matches of this pattern in the input $G$. The simplest, yet challenging, case of motif counting is when $H$ has three vertices, often called a \emph{triadic} query. 
Recent work has focused on \emph{temporal graph mining}, where the network $G$ has edges with timestamps (and directions) and $H$ has time constraints. Such networks are common representations for communication networks, citation networks, financial transactions, etc. 

Inspired by concepts in logic and database theory, we introduce the study of \emph{thresholded First Order Logic (FOL) Motif Analysis} for massive temporal networks. A typical triadic motif query asks for the existence of three vertices that form a desired temporal pattern. 
An \emph{FOL} motif query is obtained by having both existential and thresholded universal quantifiers. This allows for a query semantics that can mine richer information from networks. A typical triadic query would be "find all triples of vertices $u,v,w$ such that they form a triangle within one hour". A thresholded FOL query can express "find all pairs $u,v$ such that for half of $w$ where $(u,w)$ formed an edge, $(v,w)$ also formed an edge within an hour". 

We design the first algorithm, FOLTY, for mining thresholded FOL triadic queries. The theoretical running time of FOLTY matches the best known running time for temporal triangle counting in sparse graphs. Specifically, FOLTY runs in time  $O(m \alpha \log \sigma_{\max}).$ Here, $m$ is the number of temporal edges in the input graph, $\alpha$ is the maximum core number (degeneracy), and $\sigma_{\max}$ is the maximum edge multiplicity. We give an efficient
implementation of FOLTY using specialized temporal data structures. FOLTY has excellent empirical behavior, and can answer triadic FOL queries on graphs with nearly 70M edges is less than hour on commodity hardware. Our work has the potential to start a new research direction in the classic well-studied problem of motif analysis. 
\end{abstract}




\maketitle
\section{Introduction}

Triangle counting is a fundamental problem in network analysis, that has been extensively studied in the literature \cite{hessel2016science, kondor2021rich,jha2015path,tsourakakis2009doulion,ahmed2014graph,turk2019revisiting,pavan2013counting,seshadhri2019scalable,seshadhri2014wedge,wang2017moss,iyer2018asap,pinar2017escape, hasanetal}. This problem has found many applications in network analysis \cite{fast-gen-networks}, synthetic graph generation~\cite{SeKoPi11},
indexing graph databases \cite{graph-indexing}, and discovering communities \cite{article-network-disc, hessel2016science}.
Many real-world networks are \emph{temporal}, meaning that edges have timestamps. For instance, consider the financial network of a bank, in which the nodes represent bank accounts, and directed edges represent transfer of money between accounts~\cite{coevolve, dense-seq-streams,kovanen2011temporal}. 
Over the past few years, there has been research on defining and counting temporal motifs, which incorporate time and ordering 
constraints on edges~\cite{mackey2018chronological, paranjape2017motifs,kumar20182scent,talati2022mint,sun2019tm,gao2022scalable,kovanen2011temporal, kovanen2013temporal, talati2022ndminer}. Temporal triangle/motif counts have proven to be useful in a multitude of applications such as network classification \cite{temp-network-classification}, temporal text network analysis \cite{temp-text-analytics}, analyzing financial networks \cite{liu2023temporal, hajdu2020temporal}, etc. 


Motif (especially triangle) counting is an extremely powerful technique in graph mining. Yet, it is unable to capture richer patterns
such as the following. Suppose we want to find edges $(u,v)$ of the network where: whenever $u$ makes a transaction to $w$,
then $w$ makes a transaction to $v$ within an hour. Clearly, this is related to triangles involving $u,v,w$,
but observe that $w$ is not a fixed vertex. \emph{How can we express such richer patterns in terms of motifs?}

\subsection{First Order Logic Motifs} \label{sec:fol}

We take inspiration from the concept of First Order Logic (FOL) statements in logic and database theory.
The study of this concept is ubiquitous in database theory and parameterized graph theory~\cite{db-theory-book, fol-complexity-db-theory-1, fol-complexity-db-theory-2, fol-complexity-db-theory-3, fol-complexity-db-theory-4}, logic and designing AI expert systems \cite{FOL-Planning, FOL-Planning-2, AI-Book}. 

In the language of FOL, a motif (say, a triangle) query is an \emph{existential FOL} statement.
We are asking: do there exist vertices $u,v$ and $w$ such that the $(u,v,w)$ forms a triangle, maybe satisfying certain temporal constraints?
In mathematical language, we would write $\exists u \ \exists v \ \exists w$ $\phi(u,v,w)$, where $\phi(u,v,w)$ is a Boolean predicate
stating that there are edges among $u,v,w$ forming a triangle with the desired constraints. (A temporal motif
would enforce some timestamp conditions in $\phi(u,v,w)$.) The motif counting problem is to count
all possible solutions to the FOL statement.

A general FOL statement would involve both existential \emph{and} universal quantifiers.
For example, the FOL statement $\exists u \ \exists v \ \forall w $ $\phi(u,v,w)$ could express the 
query: do there exist vertices $u,v$ such that for all neighbors $w$ of $v$,
$(u,v,w)$ form a triangle? With a more complex predicate $\phi$, we can even express:
do there exist $u,v$ such that for all neighbors $w$ where $(u,w)$ is an edge,
$(w,v)$ is an edge that formed within 10 minutes? As we discuss later, we also
include \emph{thresholded universal quantifiers}, which represent 
 "for at least a $\tau$-fraction of neighbors" (for parameter $\tau$).

\begin{figure}
    \centering
    \begin{subfigure}{0.3\linewidth}
    \includegraphics[width=\linewidth]{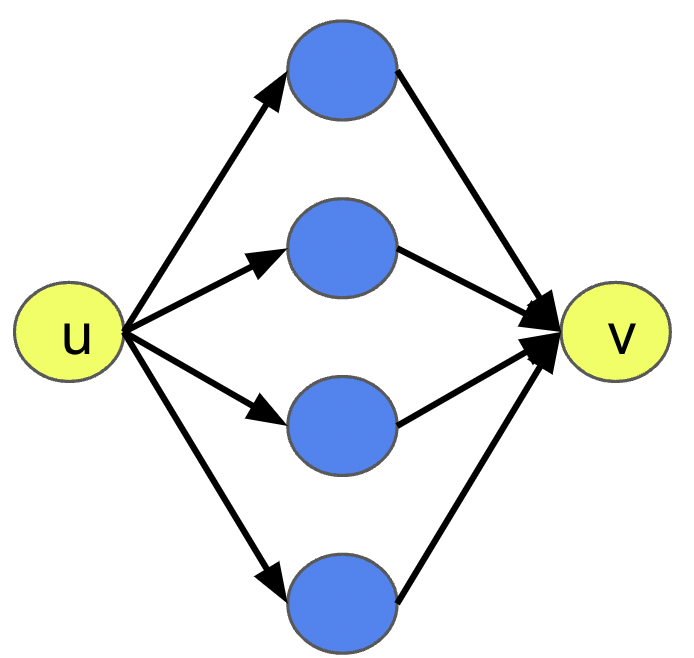}  
       \caption{Scatter-Gather}
       \label{fig:scatter-gather}
    \end{subfigure}
    \begin{subfigure}{0.4\linewidth}
   \includegraphics[width=\linewidth]{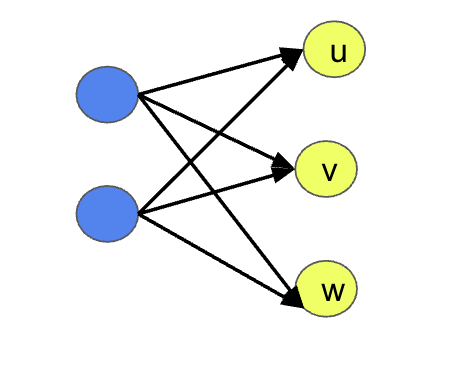}  
       \caption{Bipartite}
       \label{fig:bipartite}
    \end{subfigure}
    \begin{subfigure}{0.3\linewidth}
   \includegraphics[width=\linewidth]{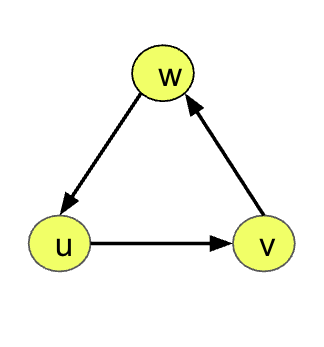}  
       \caption{Temporal Triangle for gambling}
       \label{fig:gambling-triangle}
    \end{subfigure}
    \caption{Different Money-Laundering and Gambling Patterns in Financial Transaction Networks~\cite{liu2023temporal,AlBl+23}} \label{fig:examples}
\end{figure}
There are applications for analysis of financial transaction networks.
Recent work of Liu et al.~\cite{liu2023temporal} on the Venmo transaction network
discovered that the temporal triangle in Figure-\ref{fig:examples} is a likely indication of gambling activities. 
An FOL query can determine if there are edges $(u,v)$ such that most neighbors of $w$ participate in 
such triangles. As another example, consider the two patterns, denoted "Scatter-Gather" and "Bipartite" in Figure-\ref{fig:examples}.
Altman et al. discovered these patterns as signs of money laundering in financial networks~\cite{AlBl+23}. 
Often the yellow nodes are individuals, and the blue nodes are fake or laundering accounts. The challenge
of representing these as motifs is that the number of blue nodes is unknown, so we may not know the size of the motif. 

The Scatter-Gather pattern is easily captured as a triadic FOL query: are there vertices
$u,v$ such that for (say) most $w$ where $(u,w)$ is a transaction, the transaction
$(w,v)$ occurs within (say) a few days? Observe how we apply a thresholded universal quantifier
over $w$ to capture the intermediate blue nodes, so we do not need to specify their number.
A similar FOL query captures the Bipartite money laundering pattern.

In addition, solutions to these queries can be used to discover edges, which participate in  a cohesive group of vertices in $G$. For instance, in a citation network \cite{citation-network} where nodes represent papers and edges $(u, v)$ indicate that paper $u$ cites paper $v$, we assign each edge the timestamp of $u$. Next, we run the $\exists \exists \forall$ query on this dataset for $\delta = 1$ year, list all the edges which satisfy this query, along with their common neighbors. We observe that these vertices correspond to research  papers, which were either published in the same journal or are on related topics. For instance, the temporal edge between the paper tiled "Virtual Memory,ACM Computing Surveys (CSUR)"($u$) and "Measurements of segment size"($v$) served as a solution to this query. The common neighbours of this edge were   papers titled:
\begin{enumerate}
    \item "Dynamic storage allocation systems"
    \item "Further experimental data on the behavior of programs in a paging environment"
    \item "The working set model for program behavior"
    \item "A note on storage fragmentation and program segmentation"
\end{enumerate}
Out of these 6 papers, 5 were published in "Communications of ACM", except for the paper titled "Virtual Memory", which was published in "ACM Computing Surveys". Thus, the set $\{ u\} \cup \{v \} \cup \{N(u) \cap N(v) \}$ for any solution $e = (u,v,t)$ to the $\exists \exists \forall$ query could potentially form a cohesive set of vertices.

Motivated by these applications, we ask:

{\center{\emph{Can we give a formalism for thresholded FOL motif queries on temporal networks,
and design efficient algorithms for answering such queries?}}}

\subsection{Formal Definition}\label{subsec:problem-definition}
\begin{figure}
    \centering
    \begin{subfigure}{0.4\linewidth}
       \includegraphics[width=\linewidth]{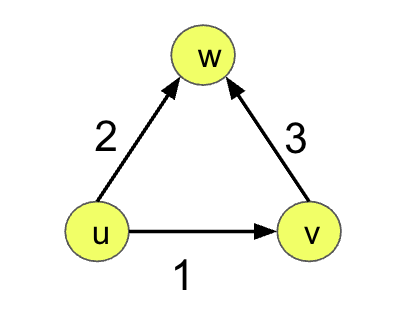} 
       \caption{Predicate for FOL queries}
       \label{fig:actual-temp-triangle}
    \end{subfigure}
    \begin{subfigure}{0.4\linewidth}
       \includegraphics[width=\linewidth]{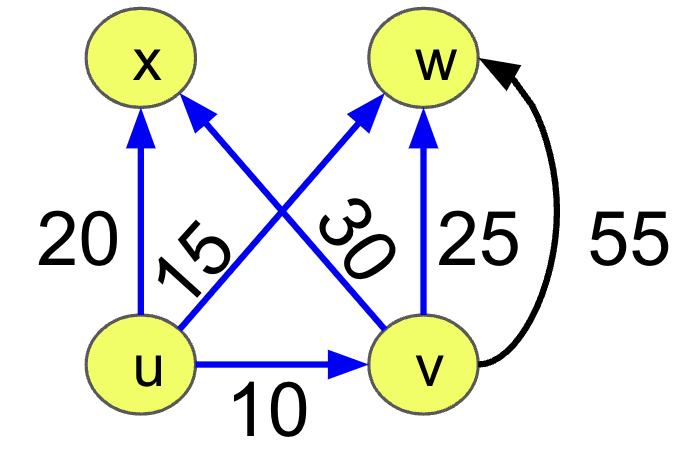}  
       \caption{Example of a solution for $\exists \exists \forall_{\tau}$ query}
       \label{fig:eea-sol}
    \end{subfigure}
    \begin{subfigure}{0.4\linewidth}
       \includegraphics[width=\linewidth]{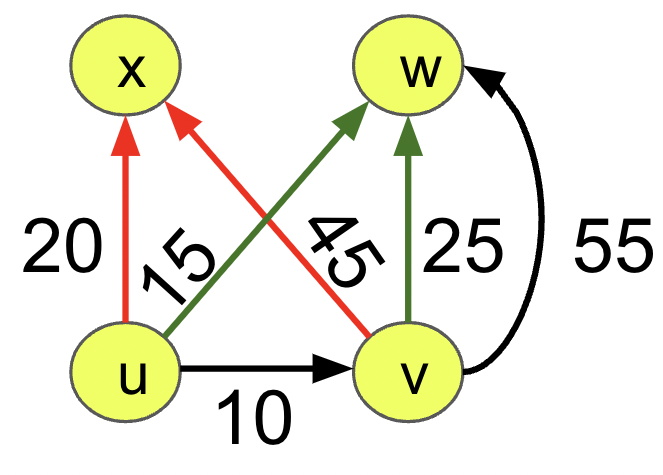}  
       \caption{Example of a non-solution for $\exists \exists \forall_{\tau}$ query}
       \label{fig:eea-non-sol}
    \end{subfigure}
%
    \caption{Consider the query $\exists u \ \exists v \ \forall w \in N(u)$, the edges $(u,v)$, $(u,w)$, and $(v,w)$ occur within 20 timesteps.
In the left figure, the edge $(u,v)$ is a solution. In the right figure, it is not a solution.
Triangle $(u,v,x)$ violations the timestamp constraint, while triangle $(u,v,w)$ violates the directions.}
       \label{fig:eea-sols}
\end{figure}

Let $G = (V,E)$ be a temporal network. Every edge in $G$ is represented as a triple $(u,v,t)$, where $u$ is the head, $v$ is the destination,
and $t$ is the edge timestamp. Note that $G$ is a multigraph, since there can be many temporal edges (in both directions)
between a pair of vertices. 
We define our main query below. We use the definitions of temporal motifs from~\cite{paranjape2017motifs,pashanasangi2021faster,pan2024accuratefastestimation}.

\begin{definition} \label{def:fol} A \emph{thresholded First Order Logic (FOL)} triadic query
is an expression $Q_1 u \ Q_2 v \ Q_3 w \ \phi(u,v,w)$ where each $Q_i$ is either
$\exists$ (existential quantifier) or $\forall_\tau$, for parameter $\tau \in (0,1)$
(thresholded universal quantifier). 

An $\exists u \ldots$ quantifier is satisfied if there exists some $u$ in the universe (vertex set)
satisfying the remaining sentence. A $\forall_\tau u$ quantifier is satisfied
if for at least a $\tau$-fraction of the universe, the remaining sentence is satisfied.
Typically, the universe for a $\forall_\tau$ quantifier is the neighborhood of a previous vertex.

The Boolean predicate $\phi(u,v,w)$ represents
a temporal motif on the vertices $u,v,w$.
A \emph{temporal motif} specifies the direction of edges in $u,v,w$, the time ordering
of edges, and an upper bound on the time interval containing these edges.
\end{definition}

To give a concrete example, consider the following query: $\exists u \ \exists v$ $\forall_{1/2} w \in N(v)$,
$(u,v,w)$ form the triangle of Figure-\ref{fig:gambling-triangle} within one hour. A solution to this query is pair $(u,v)$ of adjacent vertices, which form a triangle $(u,v,w)$ with at least half of $v$'s neighbors $w$. In each of these triangles, the edge $(u,v)$ appears first, followed by the edge $(v,w)$ and finally the edge $(w,u)$. Further,  all three edges occur within an hour of each other. What this means is that there exist temporal edges $e_{1} = (u,v,t_1)$, $e_{2} = (v,w,t_2)$ and $e_3 = (v,w,t_3)$ such that the triangle $(e_1, e_2, e_3)$ is an instance of the temporal motif $H$ specified in the predicate $\phi$. In this case, predicate  is given by the gambling triangle in Figure-\ref{fig:gambling-triangle}. Therefore, for a pair $(u,v)$ to be solution to this query, there must exist a temporal edge $e = (u,v,t)$ that forms an instance of the gambling triangle with at least half of $v$'s neighbors. In this case, we can say that the temporal edge $e = (u,v,t)$  certifies that the pair $(u,v)$ is indeed a solution to the given $\exists u \ \exists v$ $\forall_{1/2} w \in N(v)$ query. \emph{In this work, we develop algorithms, which allow us to enumerate all certificates to an $\exists \exists \forall$ query i.e. instead of just one, we find all temporal edges on $(u,v)$ that satisfy the input predicate}. We will often refer to a certificate $e = (u,v,t)$ as a solution to the $\exists \exists \forall_{\tau}$ query.

As another example, consider the following query: $\exists u \ \forall_{1/2} \ v \in N(u) \ \forall_{1/3} \ w \in N(v)$,
edge $(u,v)$ appears first, then edge $(u,w)$, and finally edge $(v,w)$, and all occur
in 40 minutes. (The predicate $\phi(u,v,w)$ is the entire sentence after the quantifiers.)
This means that we are looking for a vertex $u$ such that for at least half of its neighbors $v$,
at least one third of the neighbors $w$ of $v$ form the desired triangle specified. In particular,  for at least half of $u$'s neighbors $v$, there exists a temporal edge $e = (u,v,t)$ such that $e$ forms the desired temporal triangle  with a third of $v$'s neighbors $w$. So, for at least a one-third neighbors $w$ of $v$, $G$ contains  temporal edges $e_2 = (u,w,t_2)$ and $e_3 = (v,w,t_3)$ such that $t \leq t_2 \leq t_3 \leq t + \delta$, where $\delta$ is 40 mins.  

Observe that thresholded FOL queries provide  rich semantics for pattern mining in graphs. While we typically consider the universe for with the last quantifier as the neighborhood of the previous vertex, similar techniques can be developed for cases in which the universe happens to be, say the intersection of the neighborhoods of the previous two vertices. 

It is straightforward to define the notion of a solution to the thresholded FOL query.
Figure-\ref{fig:eea-sols} gives an example.

\begin{definition} \label{def:solution} The \emph{solution} to 
the FOL query $Q_1 u \ Q_2 v \ Q_3 w \ \phi(u,v,w)$ is the set of vertices
whose assignment to the \emph{prefix} of existential quantiers makes the statement true.

Specifically, for $\exists u \ \exists v \exists w$ queries, a solution is a triple
of vertices. For $\exists u \ \exists v \forall w$, a solution is a pair $(u,v)$.
For $\exists u \ \forall v \ \forall w$ and $\exists u \ \forall v \ \exists w$
queries, a solution is a single vertex.
\end{definition}

We do not consider queries that begin with a universal quantifier, since there is no notion
of a solution. (Using negation and de Morgan's laws, one can convert it to one of the queries
in definition-\ref{def:solution}.) Thus, we have four kinds of thresholded FOL queries,
denoted as $\exists \exists \exists$, $\exists \exists \forall$, $\exists \forall \exists$,
and $\exists \forall \forall$. The meaning of these should be clear from the above definition.

We now come to the main computational problem. Given a thresholded FOL query,
we wish to count and enumerate the number of solutions of these queries. As explained earlier, in the case
of $\exists \exists \exists$ queries, this is simply the problem of temporal
triangle counting.  In this work, we focus on \emph{Triadic Queries} i.e. FOL queries,  in which predicate is a temporal triangle. The main technical question we resolve is the following.

{\center \emph{Is there an efficient algorithm to count/enumerate solutions
of a thresholded FOL triadic query, whose theoretical running time is comparable
to the best temporal triangle counters?}}

In this paper, we present our techniques by taking the example of the predicate, which is given by the temporal triangle of Figure-\ref{fig:actual-temp-triangle}. However, our techniques can be generalized to deal with alternate temporal triangles as well.

\subsection{Contributions}\label{subsec:major-contribution}

Our main result is a theoretical algorithm, First Order Logic Triadic Yielder (FOLTY),
that can exactly compute and enumerate the number of solutions to any thresholded FOL triadic query.
As we explain below, the running time matches the best known temporal triangle counter~\cite{pashanasangi2021faster},
based on the best possible complexity known for triangle counting~\cite{chiba-nishizeki}.
We give an efficient, practical implementation of FOLTY that can process temporal
graphs with tens of millions of edges. We list out the important
contributions of our work.

\textbf{Defining thresholded FOL queries.} To the best of our knowledge, this is the first work that sets up the notion
of thresholded FOL queries for temporal networks. While the notion of FOL queries for databases has been
studied before~\cite{fol-param-1, fol-param-2}, there is no work in the data mining literature
that sets up the problems given in \ref{def:fol} and \ref{def:solution}. We believe that these definitions
set up a new research direction in motif analysis. This work also initiates the use of orientation-based methods for analyzing first-order logic (FOL) queries on attributed networks, a question which was posed in \cite{seshadhri2023tutorial}.   

    
\textbf{An algorithm with provable runtime guarantees.} FOLTY has the best possible theoretical
running time up to log factors, assuming the forty year old bound of Chiba-Nishizeki for sparse triangle counting~\cite{chiba-nishizeki}. 
We prove that
FOLTY also has a running time of $O(m\alpha \log \sigma_{\max})$,
where $m$
is the number of temporal edges, $\alpha$ is the maximum core number (degeneracy), and $\sigma_{\max}$ is the maximum
multiplicity (number of parallel edges).
This is the same running time of the best temporal triangle counters~\cite{pashanasangi2021faster}. Up to log factors, this matches
the forty year old Chiba-Nishizeki $O(m\alpha)$ bound for triangle counting~\cite{chiba-nishizeki}. 

We achieve this bound by a combination of orientation techniques of triangle counting, combined
with specially constructed temporal data structures. One of our novelties is the use of segment
trees, an efficient data structure to process intervals. FOLTY combines segment trees and degeneracy
orientations to achieve its running time. This gives the $\log \sigma_{\max}$ dependence, as opposed
to a linear dependence by enumeration methods. We note that FOLTY gets the solutions to a query \emph{without}
enumerating all temporal triangles.

    
\textbf{Unified frameworks to deal with FOL Queries.} As mentioned earlier, there are four kinds of FOL queries. We show
how to efficiently reduce $\exists \forall \forall$ and $\exists \forall \exists$ to $\exists \exists \forall$ queries. This gives a unified
algorithm for solving FOL queries. We also show how to deal with thresholded universal queries efficiently, without
enumerating entire lists of temporal edges.

\textbf{Excellent empirical performance:} We empirically evaluate the performance of FOLTY on a wide range of temporal graph datasets and various FOL queries. We conduct 
multiple experiments with varying values of thresholded quantifiers and time intervals. We observe that FOLTY works well, taking less than 20 mins. on networks with nearly 65M edges. Our algorithm for $\exists \exists \forall_{\tau}$ and $\exists \forall_{\tau_1} \forall_{\tau_2}$ consistently beats the SOTA triangle enumeration algorithm. We note that there is no previous code to solve this problem. As we discuss later, all existing algorithms either compute a total
temporal triangle count, or enumerate all temporal triangles. The number of temporal triangles can be in billions with for graphs with millions
of edges. Hence, we require the techniques developed to get FOL query solutions without a full enumeration.

\begin{figure}
    \centering
    \begin{subfigure}{0.48\linewidth}
       \includegraphics[width=\linewidth]{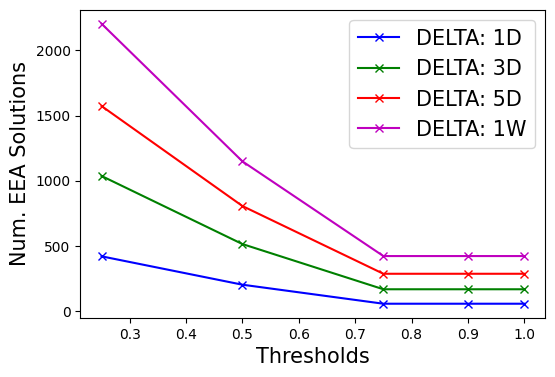}  
       \caption{$\exists \exists \forall$, for Venmo Network}
       \label{fig:eea-venmo}
    \end{subfigure}
    \begin{subfigure}{0.48\linewidth}
       \includegraphics[width=\linewidth]{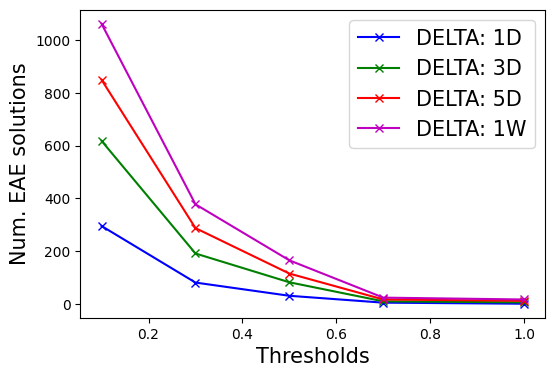}  
       \caption{$\exists \forall \exists$, for Venmo Network.}
       \label{fig:eae-venmo}
    \end{subfigure}
    \caption{We apply FOLTY to the Venmo transaction Network in~\cite{liu2023temporal}. We compute solutions for the $\exists u \ \exists v \ \forall_\tau w$ and $\exists u \ \forall_\tau v \ \exists w$,
where the triangle is $(u,v)$, $(u,w)$, $(v,w)$ (see Figure-\ref{fig:actual-temp-triangle}). The threshold $\tau$ is varied on the $x$-axis. The time interval is denoted by DELTA, and is varied to get different trends.}
    \label{fig:venmo-network}
\end{figure}

\textbf{Detecting motif trends:} We apply FOLTY to a number of real-world datasets, and get various trends over thresholded quantifiers.
An example is shown in \ref{fig:venmo-network}. For example, we get a flattening of counts as the threshold $\tau$ crosses $0.7$, meaning
that edges that form triangles with a $0.7$ fraction of their neighbors form triangles with almost their entire neighborhood.
As shown in the appendix, we can also apply FOL queries to detect edges in citation networks that are likely to be in the same topic.

%
%
\subsection{Related Work} \label{sec:related}

The subject of general motif counting is vast, so we do not attempt to cover all related work.
We refer the reader to a tutorial of Seshadhri and Tirthapura on subgraph counting~\cite{SeTi19}.
We only focus on results on temporal motif counting.

The problem of Temporal motif counting was first introduced in \cite{paranjape2017motifs}, and since then, significant progress has been made in developing efficient algorithms for this question. \cite{mackey2018chronological} presented a general-purpose algorithm that precisely counts the occurrences of any temporal motif in a graph $G$. \cite{yuan2023everest} later enhanced this algorithm to achieve better performance on GPUs. Other algorithms that provide exact motif counts include \cite{sun2019tm} and \cite{MinTemporalMotif}. Additionally, there has been considerable focus on developing specialized algorithms for mining particular motifs, such as the temporal triangle \cite{kovanen2011temporal, temp-network-classification, pashanasangi2021faster}, Butterfly-counting \cite{Butterfly-Counting} etc. 

There has also been considerable interest in developing efficient sampling methods, that generate accurate  estimates of motif counts. \cite{liu2019sampling, sarpe2021presto} divide the timespan of the input graph into multiple intervals, randomly sample one of them, and then generate exact counts within the sampled subgraph. \cite{wang2020efficient} introduced a method that samples a uniform random edge and counts the motif instances incident on it. Other sampling methods include \cite{ahmed2021online, sarpe2021oden}. More recently, \cite{pan2024accuratefastestimation} proposed an efficient approach for counting motifs on 3 and 4 vertices by introducing the concept of Path-Sampling. \cite{temporal-triangle-counting-ML-preds} utilize  predictions from ML models to estimate temporal triangle count in streaming temporal networks. 

We stress that none of these methods are geared towards solving general FOL queries. While these methods can enumerate all motifs,
it would require extra logic to deduce the solutions. Moreover, the number of temporal motifs can easily be in billions
for moderate graphs with only tens of millions of edges.

\section{Main Ideas}

\textbf{Preliminaries.} We begin with some preliminaries on the orientation methods for triangle counting.
For temporal graph $G$, the underlying static graph $G_{S}$ is constructed by removing all direction and parallel edges.
Further,  we assume that $G$ is presented to us as a list of temporal edges, sorted in the increasing order of timestamp. Given a vertex $u \in V$, $N(u) = \{v \in V$ $|$ $\{u,v\} \in E(G_{S}) \}$. Given a pair of distinct vertices $x$ and $y$, we use $E_{x,y}$ to denote the list of temporal edges from $x$ to $y$, sorted in the increasing order of their timestamp. We assume $O(1)$ access to these lists. Given $u,v \in V$, we use $\sigma(u,v)$ to denote the number of temporal edges with $u$ and $v$ as their endpoints.

Given an undirected graph $G_S$, and  an ordering $\pi$ on its vertices, we can orient the edges of $G_S$ to obtain a directed graph $G^{\rightarrow}_{\pi}$ defined as follows: $G^{\rightarrow}_{\pi} = (V, E^{\rightarrow}_{\pi})$, where $E^{\rightarrow}_{\pi} = \{(u,v)$ $|$ $\{u,v \} \in E(G)$ $\land$ $ \pi(u) < \pi(v)\}.$ Given $u \in V$, define $N^{+}_{\pi}(u)$ as the out-neighbourhood of $u$ in $G^{\rightarrow}_{\pi}$ and $N^{-}_{\pi}(u)$ as its in-neighbourhood. Next, we introduce the notion of degeneracy of $G_S$.

\begin{definition}\label{def:degeneracy}
    The \emph{degeneracy} of an undirected graph $G$ is the smallest integer $\alpha$, such that there exists an ordering of $V$ with $|N^{+}_{\pi}(v)| \leq \alpha$ for every $v \in V$. The resulting ordering is called \emph{degeneracy ordering}.
\end{definition}

The degeneracy ordering $\pi$ of an undirected graph $G$ can be obtained by repeatedly removing  the lowest degree vertex from $G$ \cite{Matula-Beck}. If two or more vertices have the same lowest degree, then the ties are resolved based on their id. Further, the order in which the vertices are removed gives the degeneracy ordering. The degeneracy ordering of $G$ can be computed in $O(|E(G)|)$ time.

Given an edge $e = \{u,v\}$ in $G_{S}$, its degree $d_{e}$  is defined as the degree of its lower-degree endpoint. A classic result of Chiba-Nishizeki states that
\begin{lemma}\label{lemma:chiba-nishzeki}\cite{chiba-nishizeki}
    $\sum_{e \in E(G)}d_{e} = O(m \alpha)$
\end{lemma}
For a temporal edge $e = (u,v,t)$, we call $u$ as the \textit{source} of this edge if $\pi(u) < \pi(v)$, else the source will be $v$.

\textbf{How to solve FOL queries.} Consider the predicate given by the temporal triangle in Figure-\ref{fig:actual-temp-triangle}. A triple $(u,v,w)$ of vertices forms  a $\delta$-temporal triangle if there exist temporal edges $(u,v,t_{1}), (u,w,t_{2})$ and $(v,w,t_{3})$ such that $t_{1} \leq t_{2} \leq t_{3} \leq t_{1} + \delta$ (more concretely, these temporal edges should satisfy the constraints specified in the input query). Similarly,  a temporal edge $e = (u,v,t)$ forms a $\delta$-temporal triangle with a vertex $w$ if there exist temporal edges $(u,w,t_{2})$ and $(v,w,t_{3})$ which satisfy $t \leq t_{2} \leq t_{3} \leq t + \delta$. Clearly, these edges form a triangle that matches the motif specified  in the input predicate (refer  Figure-\ref{fig:actual-temp-triangle}).

Now let us consider the $\exists \exists \forall_\tau$ query with a  threshold $\tau \in (0,1)$. A temporal edge $e = (u,v,t)$ is a solution to this query if at least $\tau$ fraction of $v$'s neighbors form at least one $\delta$-temporal triangle with $e$. 

Given a temporal network $G$, the first step is to construct  its underlying static graph $G_{S}$ and use it to obtain the directed graph $G^{\rightarrow}_{\pi}$, where $\pi$ is the degeneracy ordering of $G_{S}$. Our algorithm  outputs 2 arrays of size $m$, namely \emph{in-count} and \emph{out-count}. These arrays are indexed by the temporal edges in $G$. For each  temporal edge $e$, in-count[$e$] stores the number of vertices in $N^{-}_{\pi}(e_s)$ which form at least one $\delta$-temporal triangle with $e$. Here $e_s$ is used to denote the source  of $e$. A similar notion can be defined for out-count[$e$]. 

Note that every neighbor of $v$, that forms a $\delta$-temporal triangle with $e = (u,v,t)$, must be a common neighbor of both $u$ and $v$. Therefore, every such vertex will be in the neighborhood of $e$'s source. Suppose $k$  neighbors of $e_s$  participate in a $\delta$-temporal triangle with $e$. Then, it follows that exactly $k$ neighbors of $v$ form a $\delta$-temporal triangle with $e$. As a result,  if in-count[$e$] + out-count[$e$] $\geq \tau |N(v)|$, then  at least $\tau$ fraction of $v$'s neighbors form the desired triangle with $e$. In this case, $e$ will be a solution to the $\exists \exists \forall_{\tau}$ query.

The two arrays can also be used to determine solutions for the $\exists \forall_{\tau_1} \forall_{\tau_2}$ and $\exists \forall_{\tau} \exists$ queries. For the former query, we  look for vertices $v$, such that for at least $\tau_{1}$ fraction of $v$'s neighbors $x$, there exists a temporal edge $e = (v,x,t)$, such $\tau_{2}$ fraction of $x$'s neighbors form a $\delta$-temporal triangle with $e$. Observe that for a given vertex $v$, and for every such neighbor $x$ of $v$, the temporal edge $e = (v,x,t)$ must be a solution to the $\exists \exists \forall_{\tau_2}$ query. Consequently, solutions to the given $\exists \forall_{\tau_{1}} \forall_{\tau_2}$ query can be obtained  by enumerating solutions to the $\exists \exists \forall_{\tau_2}$ query.  

A similar approach can be used to enumerate the solutions to the $\exists \forall_{\tau} \exists$ query. A vertex $v$ is a solution to this  query if, for at least $\tau$-fraction of $v$'s neighbors $x$, there exists a temporal edge $e = (v,x,t)$ such that $e$  is incident on at least one $\delta$-temporal triangle. In particular, for at least $\tau |N(v)|$ neighbors $x$ of $v$, $G$ must contain a temporal edge $e = (v,x,t)$ for which in-count[$e$] + out-count[$e$] $\geq 1$. From this discussion, it is evident that the arrays in-count[$e$] and out-count[$e$] are crucial for determining solutions for all the three queries.

\textbf{Challenges.} \emph{First Attempt:} Let us consider a simple approach which, at very high level, begins by enumerating all the triangles in $G_{S}$. Each temporal edge is assigned a counter which is initially 0. For every temporal edge $e = (u,v,t)$, its associated counter $c_e$ is used to store the number of vertices in $N(u) \cap N(v)$, which form at least one $\delta$-temporal triangle with $e$. If $c_{e} \geq \tau |N(v)|$, then $e$ will be a solution to  $\exists \exists \forall_{\tau}$ query. To compute this value, we simply iterate over all the triangles $(u,v,w)$ in $G_{S}$.  For every triangle $(u,v,w)$, we fetch the temporal edge lists $E_{u,v}, E_{u,w}$ and $E_{v,w}$.  These are potential matches to the edges labeled 1,2 and 3 in the motif, given in Figure-\ref{fig:actual-temp-triangle}.  Our goal is identify the edges in $E_{u,v}$, that form a $\delta$-temporal triangle with $w$. To do so, for every temporal edge $e = (u,v,t) \in E_{u,v}$, we look for a pair of edges $(e_2 = (u,w,t_2), e_3 = (v,w,t_3)) \in E_{u,w} \times E_{v,w}$ that satisfies $t \leq t_2 \leq t_3 \leq t + \delta$.  Note that the existence of such a pair immediately implies that $w$ forms a $\delta$-temporal triangle with $e$. In this case, we increment the counter $c_{e}$ by 1. Since these edge lists are sorted in the increasing order of their timestamp, we can find such a pair (if it exists) for every temporal edge  in $E_{u,v}$ in $O(\sigma(u,v) + \sigma(u,w) + \sigma(v,w))$ time. This procedure is also repeated for the temporal edges going from $v$ to $u$.  

Now, let us analyze the time complexity of this algorithm. Triangle enumeration in $G_{S}$ can be done in $O(m \alpha)$ time. The time spent in updating the values of the counters $c_{e}$ is upperbounded by $c\sum_{\{u,v\} \in E(G_{S})}\sum_{w \in N(u) \cap N(v)}(\sigma(u,v) + \sigma(u,w) + \sigma(v,w))$ for some suitable constant $c$. Now observe that the number of times $\sigma(u,v)$ appears in this sum is equal to the number of triangles incident on the edge $\{u,v\}$ in $G_{S}$. This quantity  is upperbounded by $d_{u,v}$ i.e. the degree of the edge $\{u,v\}$. Therefore, the previous sum is at most c$\sum_{(u,v) \in E(G_{S})}d_{u,v}\sigma(u,v) \leq d_{\max}\sum_{(u,v) \in E(G_{S})}\sigma(u,v) = c\cdot md_{\max}$, where $d_{\max} = \max_{e \in E(G_{S})}d_e$. Note that $d_{\max}$ can be must larger than $\alpha$. Clearly, this approach does not work.

\emph{Challenge 1:}  Lets consider an alternative approach. For a given temporal edge $e = (u,v,t)$, we will use $s_{e}$ to denote its source and $y_e$ to denote its other endpoint. Recall that $s_e =u$ iff $\pi(u) < \pi(v)$. In the previous case, in order to determine the value of the counter $c_{e = (u,v,t)}$, we would iterate over all triangles $(u,v,w)$ that contain the edge $\{u,v \}$. For every such triangle, we spend $O(\sigma(u,v) + \sigma(u,w) + \sigma(v,w))$ time in figuring out which the temporal edges in $E_{u,v}$  form a $\delta$-temporal triangle with $w$. Now assume that there exists a procedure, which  
\begin{enumerate}
    \item computes the correct value of $c_{e}$ by looking solely at the vertices in $w \in N^{+}_{\pi}(s_{e}) \cap N(y_{e})$ (as opposed to $N(s_e) \cap N(y_e)$), where $\pi$ is the degeneracy ordering of $G_{S}$ and
    \item  determines which edges in $E_{u,v}$ are incident on a $\delta$-temporal triangle with $w$ in $O(\sigma(s_e, y_e) + \sigma(s_e, w))$ time (i.e. by making a pass over the temporal edges incident on the source vertex only ($E_{s_e,y_e}, E_{s_e, w}$) , as opposed to iterating through all the three  edge-lists $E_{s_e,y_e}, E_{s_e, w}$ and $E_{y_e,w}$). 
\end{enumerate}
In this case, given an edge $\{u,v\}$, instead of iterating over all of $N(u) \cap N(v)$, it suffices to look into $N^{+}_{\pi}(s_e ) \cap N(y_e)$. Recall that if $(u,v) \in E(G^{\rightarrow}_{\pi}),$ then $\pi(u)$ must be less that $\pi(v)$ (i.e. $u$ is the source vertex). Therefore, for every $(u,v) \in E(G^{\rightarrow}_{\pi})$ and $w \in N^{+}_{\pi}(u)$, if $w \in N(v)$ then we fetch the edge lists $E_{u,v}, E_{u,w}, E_{v,w}$ and execute the above on these lists. Note that since $u$ is the source, the algorithm can iterate only through the temporal edges incident on the source $u$. These are precisely the edges in  lists $E_{u,v}$ and $E_{u,w}$. Similarly, when we fetch the edge lists $E_{v,u}, E_{v,w}$ and $E_{u,w}$(potential choices for edges labeled 1,2 and 3 in Figure-\ref{fig:actual-temp-triangle}), the algorithm can iterate through the temporal edges in the lists $E_{u,w}$ and $E_{v,u}$ only, since they are the ones incident on the source $u$. As a result, overall time complexity of the algorithm can be upper bounded by  $c\sum_{(u,v) \in E(G_{S})}\sum_{w \in N^{+}{\pi}(s_{e}) \cap N(y_{e})}(\sigma(u,v) + \sigma(u,w))$. Now observe that a single term $\sigma(u,v)$ appears at most $|N^{+}_{\pi}(s_e)|$ times in this sum, which is upper bounded by $\alpha$. This leads to the $O(m \alpha)$ running time. Hence, the first challenge is to \emph{come up with a procedure that satisfies 1 and 2}

\emph{Challenge 2:} For an edge $e = \{u,v\}$, $N(u) \cap N(v)$ can be partitioned into $N^{+}_{\pi}(s_e) \cap N(y_e)$ and $N^{-}_{\pi}(s_e) \cap N(y_e)$. Given a temporal edge $e = (u,v,t)$, in order to compute the value of the counter $c_e$, we need to know which vertices in $N^{+}_{\pi}(s_e)$ and $N^{-}_{\pi}(s_e)$ form a $\delta$-temporal triangle with $e$. Turns out, it is relatively easy to identify the out-neighbors in $N^{+}_{\pi}(s_e)$, which form a $\delta$-temporal triangle with $e$ . The main challenge is dealing with the in-neighbors $N^{-}_{\pi}(s_e)$. Consider $w \in N^{-}_{\pi}(s_e)$. This is equivalent to saying that $s_e \in N^{+}_{\pi}(w)$. Further, if $w \in N(y_{e})$, then $y_e \in N^{+}_{\pi}(w)$ since $\pi(s_e)< \pi(y_e)$. Since the algorithm iterates over every $(x,y ) \in E(G^{\rightarrow}_{\pi})$ and $z \in N^{+}_{\pi}(x)$, the contribution of $w$ to the value of $c_{e}$ will be accounted when considering the edge $(w, s_{e}) \in E(G^{\rightarrow}_{\pi})$ and $v \in N^{+}_{\pi}(w)$ i.e. when $w$ is the source vertex.  Our goal is find, for each edge in $E_{u,v}$, a pair of edges in $E_{u,w} \times E_{v,w}$ that forms a $\delta$-temporal triangle with $w$. However, note that we are no longer allowed to iterate through the edges in the list $E_{u,v}$. We can only go over  the edges in the lists $E_{w,v}$ and $E_{w,u}$ since the source vertex is $w$. This is the second challenge, where we would like to determine the edges in $E_{u,v}$ that form a $\delta$-temporal triangle with $w$, without going over the edges in the list $E_{u,v}$ for every $w \in N^{-}_{\pi}(s_e)$. 
\section{$\exists \exists \forall$ Query}
Given a temporal graph $G = (V,E)$, our goal is to find solutions to the $\exists \exists \forall$ query mentioned in \ref{def:solution}.




    The algorithm begins by constructing the underlying static graph  $G_{S}$. This can be done by making a single pass through the edges of $G$. Next, for every static edge $\{ u,v\} \in G_{S}$, it defines a counter $c_{uv}$. The counter $c_{uv}$ is used to store the number of common neighbours of $u$ and $v$ in $G_{S}$. In order to compute the value of $c_{uv}$, the algorithm first determines the lower degree endpoint of the static edge $\{u,v\}$. Assume $u$ to be this endpoint. Next, it steps through the neighbours of $u$ in $G_{S}$. For every $w \in N(u)$, if $w$ is adjacent to $v$, it increases the value of $c_{uv}$ by 1. Finally, the algorithm repeats this procedure for every static edge $\{u,v\} \in G_{S}$.

Next, it computes the degeneracy ordering $\pi$ of $G_{S}$, followed by the construction of the directed graph $G^{\rightarrow}_{\pi}$. Our goal is to compute in-count[$e$] and out-count[$e$] for every temporal edge $e$. 

 Broadly speaking, it is relatively easier to determine the value of out-count[$e$]. However, similar approach cannot be used to compute in-count[$e$]. This is primarily because in $G^{\rightarrow}_{\pi}$, the out-degree of each vertex is upper bounded by the degeneracy $\alpha$, however the in-degree can be arbitrarily large.  This leads to different ways of dealing with the in-neighbours and out-neighbours, each of which are presented in sections \ref{subsec:out-ngbrs-eefa} and \ref{subsec:in-ngbrs-eefa} respectively.
\subsection{Dealing with Out-Neighbours}\label{subsec:out-ngbrs-eefa}
Given an edge $(u,v) \in G^{\rightarrow}_{\pi}$, the goal is to compute out-count[$e$] for every temporal edge $e = (u,v,t)$ (and $(v,u,t)$). We defer the proofs of the theorems in this section to  the supplementary material. Note that if $(u,v) \in G^{\rightarrow}_{\pi}$, then $\pi(u) < \pi(v)$.

 The algorithms presented in this section use the following subroutines as primitives:
\begin{asparaenum}
    \item \textsc{FindExceedingEntryLS}$(L_{1}, L_{2}) : $ $L_{1}$ and $L_{2}$ are lists of temporal edges, sorted in the increasing order of timestamp. The function returns a list $L_{12}$ of size $|L_{1}|$. For every $e \in L_{1}$, $L_{12}[e]$ stores the index of the first edge in $L_{2}$ with timestamp at least $t(e)$. Further, it does so in time $O(|L_{1}| + |L_{2}|)$.
\item \textsc{FindExceedingEntryBS}$(L_{1}, L_{2}) : $ $L_{1}$ and $L_{2}$ are lists of temporal edges, sorted in the increasing order of timestamp. The function returns a list $L_{12}$ of size $|L_{1}|$. For every $e \in L_{1}$, the function uses binary search to determine the index of the first edge in $L_{2}$ whose timestamp exceeds $t(e)$. Thus the overall running time of the algorithm is $O(|L_{1}|\log |L_{2}|).$
\item \textsc{FindBoundingEntry}$(L_{1}, L_{2},y) : $ As before, $L_{1}$ and $L_{2}$ are lists of temporal edges, sorted in the increasing order of timestamp. The function returns a list $L_{12}$ of size $|L_{1}|$. For every $e \in L_{1}$, $L_{12}[e]$ stores the index of the last edge in $L_{2}$ with timestamp at most $t(e) + y$. This function runs in $O(|L_{1}| + |L_{2}|)$ time.
\end{asparaenum}

\smallskip

Let us fix an edge $(u,v)$ in $G^{\rightarrow}_{\pi}$ and an out-neighbour $w$ of $u$. Assume that $v$ is adjacent to $w$ in $G_{S}$. If not, then it is obvious that $w$ cannot form a triangle with $e$. Suppose $L_{1} \gets E_{u,v}$. The goal is to compute the value of out-count[$e$] for every $e \in L_{1}$. Set $L_{2} \text{ to } E_{u,w}$ and $L_{3} \text{ to } E_{v,w}$. An edge $e \in L_{1}$ forms at least one $\delta$ temporal triangle with $w$, if there exists a pair $(e_{2}, e_{3}) \in L_{2} \times L_{3}$ which satisfies $t(e) \leq t(e_{2}) \leq t(e_{3}) \leq t(e) + \delta$. Thus, for an edge $e \in L_{1}$, if such a pair exists in $L_{2} \times L_{3}$, then the algorithm increments out-count[$e$] by 1. This leads to the following natural question: "How to find such a pair?"

 Given $L_{1}, L_{2}$ and $L_{3}$,  the algorithm proceeds as follows (refer algorithm-\ref{algo:eef-outngbrs-case1}): 
\begin{asparaenum}
    \item It invokes \textsc{FindExceedingEntryLS}$(L_{1}, L_{2})$ to get the list $L_{12}$. We will abuse  notation and use $L_{12}[e]$ to  denote the index, as well the edge at this index in $L_{2}$.
    \item Next, it invokes \textsc{FindExceedingEntryBS}$(L_{2}, L_{3})$ to obtain the list $L_{23}$. We will abuse  notation and use $L_{23}[e]$ to  denote the index, as well the edge at this index in $L_{3}$.
    \item Finally, for every $e \in L_{1}$,  it checks whether $t(e) \leq t(L_{12}[e]) \leq t(L_{23}[L_{12}[e]]) \leq t(e) + \delta$. If this condition holds, then it increments out-count[$e$] by 1. Otherwise, it proceeds to the next edge of $L_{1}$.
\end{asparaenum}

\smallskip

The following theorem shows that $e$ forms a $\delta$-temporal triangle with $w$ if and only if $(e, L_{12}[e], L_{23}[L_{12}[e]])$ form a $\delta$-temporal triangle.

\begin{theorem}\label{thm:out-ngbrs-case-1}
     Consider a temporal edge $e = (u,v,t)$ with $\pi(u) < \pi(v)$ and an out-neighbour $w$ of $u$ which is adjacent to $v$. $e$  forms at least one $\delta$-temporal triangle with $w$ if and only if $t(e) \leq t(L_{12}[e]) \leq t(L_{23}[L_{12}[e]]) \leq t(e) + \delta$.
 \end{theorem}
 \begin{proof}\label{pf:out-ngbrs-case-1}
     If $t(e) \leq t(L_{12}[e]) \leq t(L_{23}[L_{12}[e]]) \leq t(e) + \delta$ holds, then the edges $e$ in $L_{1}$, $L_{12}[e]$ in $L_{2}$ and $L_{23}[L_{12}[e]]$ in $L_{3}$  form a $\delta$-temporal triangle.

     Now suppose that the edges $e_{1}, e_{2}$ and $e_{3}$ form a $\delta$-temporal triangle, with $e_{i} \in L_{i}$ for $i = 1,2,3$ and $e_{1} = e$. It must be the case that $t(e) \leq t(e_{2}) \leq t(e_{3}) \leq t(e) + \delta$. Based on the definition of the lists $L_{12}$ and $L_{23}$, we know that $t(e) \leq t(L_{12}[e]) \leq t(e_{2})$, $t(e_{2}) \leq t(L_{23}[e_{2}]) \leq t(e_{3})$ and $t(L_{12}[e]) \leq t(L_{23}[L_{12}[e]])$. Since $t(L_{12}[e]) \leq t(e_{2})$, it follows that $t(L_{23}[L_{12}[e]]) \leq t(L_{23}[e_{2}])$. Putting it all together, we get $t(e) \leq t(L_{12}[e]) \leq t(L_{23}[L_{12}[e]]) \leq t(L_{23}[e_{2}]) \leq t(e_{3}) \leq t(e) + \delta$.  Therefore $e, L_{12}[e]$ and $L_{23}[L_{12}[e]]$ form a $\delta$-temporal triangle. This completes the proof.
 \end{proof}
 
 

\begin{algorithm}[t]
\caption{: \textsc{OutNeighboursCase1}($G, G_{S},u,v,w, \delta$, out-count[])}
\label{algo:eef-outngbrs-case1}
\begin{flushleft}
\textbf{Input}: $G, G^{\rightarrow}_{\pi}, G_{S}, \delta$  \\
\begin{algorithmic}[1]

\State $L_{1} \gets E_{u,v}, L_{2} \gets E_{u,w}, L_{3} \gets E_{v,w}$
\State $L_{12} \gets$  \textsc{FindExceedingEntryLS}$(L_{1}, L_{2})$\
\State $L_{23} \gets$  \textsc{FindExceedingEntryBS}$(L_{2}, L_{3})$\
\For{$e = (u,v,t) \in L_{1}$}
\If{$t(e) \leq t(L_{12}[e]) \leq t(L_{23}[L_{12}[e]] \leq t(e) + \delta$}
\State out-count[$e$] $\gets$ out-count[$e$] + 1
\EndIf
\EndFor
\end{algorithmic}
\end{flushleft}
\end{algorithm}

\begin{algorithm}[t]
\caption{: \textsc{OutNeighboursCase2}($G, G_{S},u,v,w, \delta$, out-count[])}
\label{algo:eef-outngbrs-case2}
\begin{flushleft}
\textbf{Input}: $G, G^{\rightarrow}_{\pi}, G_{S}, \delta$  \\
\begin{algorithmic}[1]
\State Get the lists $L_{1} \gets E_{v,u}, L_{2} \gets E_{v,w}, L_{3} \gets E_{u,w}$
\State $L_{13} \gets$  \textsc{FindBoundingEntry}$(L_{1}, L_{3}, \delta)$\
\State $L_{12} \gets$  \textsc{FindExceedingEntryBS}$(L_{1}, L_{2})$\
\For{$e = (v,u,t) \in L_{1}$}
\If{$t(e) \leq t(L_{12}[e]) \leq t(L_{13}[e]) \leq t(e) + \delta$}
\State out-count[$e$] $\gets$ out-count[$e$] + 1
\EndIf
\EndFor
\end{algorithmic}
\end{flushleft}
\end{algorithm}

\begin{algorithm}[t]
\caption{: \textsc{OutNeighbours}($G, G^{\rightarrow}_{\pi}, G_{S},\delta$, out-count[])}
\label{algo:eef-outngbrs}
\begin{flushleft}
\textbf{Input}: $G, G^{\rightarrow}_{\pi}, G_{S}, \delta$  \\
\begin{algorithmic}[1]
\For{$(u,v) \in E(G^{\rightarrow}_{\pi})$}
\For{$w \in N^{+}(u)$}
\If{$w$ is adjacent to $v$}
\State \textsc{OutNeighboursCase1}($G, G_{S},u,v,w, \delta$, out-count[])\
\State \textsc{OutNeighboursCase2}($G, G_{S},u,v,w, \delta$, out-count[])\
\EndIf
\EndFor
\EndFor
\end{algorithmic}
\end{flushleft}
\end{algorithm}

Next, consider the case in which $L_{1}$ consists of the temporal edges from $v$ to $u$ i.e. $L_{1} \gets E_{v,u}$. Set $L_{2} \text{ to } E_{v,w}$ and $L_{3} \text{ to } E_{u,w}$. As before, an edge $e$ in $L_{1}$  forms a $\delta$-temporal triangle with $w$ if there exists a pair in $L_{2} \times L_{3}$, that satisfies the temporal window and ordering constraints. Since the edges in $L_{2}$ are not incident on the source $u$ of the edge $(u,v)$, the algorithm cannot make a pass through the edges in this list. Therefore, it uses uses algorithm-\ref{algo:eef-outngbrs-case2} instead.
The following claim establishes the correctness of algorithm-\ref{algo:eef-outngbrs-case2}.
 
\begin{theorem}\label{thm:out-ngbrs-case-2}
     Consider a temporal edge $e = (v,u,t)$ with $\pi(u) < \pi(v)$ and an out-neighbour $w$ of $u$ which is adjacent to $v$ in $G_{S}$. $e$ will form at least one $\delta$ temporal triangle with $w$ if and only if $t(e) \leq t(L_{12}[e]) \leq t(L_{13}[e]) \leq t(e) + \delta$, where the lists $L_{1}, L_{2}, L_{3}, L_{12}$ and $L_{23}$ are defined as above.
\end{theorem}
\begin{proof}
    If the condition $t(e) \leq t(L_{12}[e]) \leq t(L_{13}[e]) \leq t(e) + \delta$ is satisfied, then the edges $e$ in $L_{1}$, $L_{12}[e]$ in $L_{2}$ and $L_{13}[e]$ in $L_{3}$  form a $\delta$-temporal triangle.

     Now suppose that the edges $e_{1}, e_{2}$ and $e_{3}$ form a $\delta$-temporal triangle, with $e_{i} \in L_{i}$ for $i = 1,2,3$ and $e_{1} = e$. Clearly $t(e) \leq t(e_{2}) \leq t(e_{3}) \leq t(e) + \delta$. Based on the definitions of the lists $L_{12}$ and $L_{13}$, it follows that $t(e) \leq t(L_{12}[e]) \leq t(e_{2})$ and $t(e_{3}) \leq t(L_{13}[e]) \leq t(e) + \delta$. Putting these together, we get $t(e) \leq t(L_{12}[e]) \leq t(e_{2}) \leq t(e_{3}) \leq t(L_{13}[e]) \leq t(e) + \delta$. Therefore, we can conclude that the edges $e, L_{12}[e]$ and $L_{13}[e]$ form a $\delta$-temporal triangle.
\end{proof}
Combining the two cases, the overall procedure for dealing with the out-neighbours is outlined in algorithm-\ref{algo:eef-outngbrs}. 
\begin{theorem}\label{thm:run-time-out-ngbrs-eea}
    The procedure \textsc{OutNeighbours} runs in $O(m\alpha \log \sigma_{\max})$ time.
\end{theorem}
\begin{proof}\label{pf:run-time-out-ngbrs-eea}
The overall time spent by the algorithm is \\
$O(\sum_{(u,v) \in G^{\rightarrow}_{\pi}}\sum_{w \in N^{+}_{\pi}(u)}(\sigma(u,v) + \sigma(u,w))\log \sigma_{\max})$. Since $|N^{+}_{\pi}(u)| \leq \alpha$ for every $u \in V(G^{\rightarrow}_{\pi})$, it follows that the time complexity can be upper bounded by $O(\sum_{(u,v) \in G^{\rightarrow}_\pi}(\sigma(u,v))(\alpha \log \sigma_{\max})) = \\
O(m \alpha \log \sigma_{\max})$.
\end{proof}
 Note that in the end, for every $e = (u,v,t)/(v,u,t) \in E(G)$, assuming $\pi(u) < \pi(v)$, out-count[$e$] will contain the number of out-neighbours of $u$ which  form at least one $\delta$-temporal triangle with $e$.

\subsection{Dealing with In-Neighbours}\label{subsec:in-ngbrs-eefa}
After dealing with the out-neighbours, we need to compute in-count[$e$] for every temporal edge $e$.

Fix a temporal edge $e = (u,v,t)$. Assume for the moment that $\pi(u) < \pi(v)$. Suppose $w \in N^{-}_{\pi}(u) \cap N(v)$. Observe that if $w \in N^{-}_{\pi}(u) \cap N(v)$, then $u \in N^{+}_{\pi}(w)$. Also $v \in N^{+}_{\pi}(w)$ as $\pi(u) < \pi(v)$. Consequently, in-count[$e$]  must be updated when considering the edge $(w,u) \in E(G^{\rightarrow}_{\pi})$ and its out-neighbor $v \in N^{+}_{\pi}(u)$ .  In this case, we set $L_{2}$ to $E_{u,w}$ and $L_{3}$ to $E_{v,w}$ respectively. Since $w$ is the source, we cannot enumerate the edges in the list $E_{u,v}$.

As before, $e$ forms a $\delta$-temporal triangle with $w$ if there is a pair of edges $(e_{2}, e_{3}) \in L_{2} \times L_{3}$ that satisfies $t \leq t(e_{2}) \leq t(e_{3}) \leq t + \delta$.   For every $e \in L_{2}$, let $L_{23}[e]$ be the first edge in $L_{3}$ with timestamp at least $t(e)$. Define $I_{w,u,v}$ as the following collection of intervals,  $I_{w,u,v} := \{ [\max \{0, t(L_{23}[e]) - \delta \}, t(e)]$ $|$ $\{e \in L_{2}\}$ $\land$ $\{t(e) \leq t(L_{23}[e]) \leq t(e) + \delta\} \}$. We begin by stating the following theorem.

\begin{theorem}\label{thm:in-ngbrs-intervals}
    The temporal edge $e = (u,v,t)$ forms at least one $\delta$-temporal triangle with $w$ if and only if there exists an interval in $I_{w,u,v}$ that contains $t = t(e)$.
\end{theorem}
\begin{proof}\label{pf:in-ngbrs-intervals}
    Let us assume that the edges $e_{1}, e_{2}$ and $e_{3}$ form a $\delta$-satisfying triangle, where $e_{1} = e$ and $e_{i} \in L_{i}$ for $i = 1,2,3$. We would like to show that there exists an interval in $I_{w,u,v}$ which contains $t(e_{1}).$ From the definition of the list $L_{23}$, we know that $t(L_{23}[e_{2}]) \leq t(e_{3})$. Therefore, $t(L_{23}[e]) - \delta \leq t(e_{3}) - \delta$. Since $e_{1}, e_{2}$ and $e_{3}$ form a $\delta$-temporal triangle,  $t(e_{3}) \leq t(e_{1}) + \delta \leq t(e_{2}) + \delta$. As a consequence, $t(e_{3}) - \delta \leq t(e_{2})$. Hence we get $t(L_{23}[e_{2}]) - \delta \leq t(e_{3}) - \delta \leq t(e_{2})$. Equivalently, $t(e_{3}) - \delta \in [t(L_{23}[e_{2}]) - \delta, t(e_{2})]$. Since $t(e_{3}) \leq t(e_{1}) + \delta$, it follows that $t(e_{3}) - \delta \leq t(e_{1}) \leq t(e_{2})$. Since $t(e_{3}) - \delta \in [t(L_{23}[e_{2}]) - \delta, t(e_{2})]$, we can conclude that $t(e_{1}) \in [t(L_{23}[e_{2}]) - \delta, t(e_{2})].$ Further, since  $t(e_{2}) \leq t(L_{23}[e_{2}]) \leq t(e_{3}) \leq t(e_{1}) + \delta \leq t(e_{2}) + \delta$, we can conclude that $t(e_{2}) \leq t(L_{23}[e]) \leq t(e_{2}) + \delta$. Combined with the fact that $e_{2} \in L_{2}$ implies that the interval $[t(L_{23}[e_{2}]) - \delta, t(e_{2})] \in I_{w,u,v}$. Hence we have found an interval in $I_{w,u,v}$ that contains $t(e_{1})$.

    For the other direction, let us assume that there exists an interval $I = [t_{1}, t_{2}]$ in $I_{w,u,v}$ that contains $t = t(e)$.  Then there must exist an edge $e^{'} \in L_{2}$ such that $t(L_{23}[e^{'}]) - \delta = t_{1}$ and $t(e^{'}) = t_{2}$. Therefore, $t(L_{23}[e^{'}]) - \delta \leq t(e) \leq t(e^{'})$. Since $t(e^{'}) \leq t(L_{23}[e^{'}])$, we have that  $t(L_{23}[e^{'}]) - \delta \leq t(e) \leq t(e^{'}) \leq t(L_{23}[e^{'}])$, which is equivalent to $t(e) \leq t(e^{'}) \leq t(L_{23}[e^{'}]) \leq t(e) + \delta$. So there exist edges $e^{'} \in L_{2}$ and $L_{23}[e^{'}] \in L_{3}$ that form a $\delta$-satisfying triangle with the edge $e$.

    This completes the proof.
\end{proof}
 Note that a similar result can be proved for the temporal edges of the form $(v,u,t)$ by setting $L_{1} \gets E_{v,u}, L_{2} \gets E_{v,w}$ and $L_{3} \gets E_{u,w}$. Equipped with theorem-\ref{thm:in-ngbrs-intervals},  the algorithm proceeds as follows:  for every edge $(u,v) \in G^{\rightarrow}_{\pi}$ and $w \in N^{+}_{\pi}(u) \cap N(v)$, it gets the edge-lists $L_1 \gets E_{w,v}, L_{2} \gets E_{w,u}$ and $L_{3} \gets E_{v,u}$. The goal is compute in-count[$e$] for every $e \in L_{1}$. In the following discussion, we will stick to the case $L_1 \gets E_{w,v}, L_{2} \gets E_{w,u}, L_{3} \gets E_{v,u}$, since the other one can be handled analogously. 

Once the algorithm has the lists $L_{2}$ and $L_{3}$, it invokes \\ \textsc{FindExceedingEntryLS($L_{2}, L_{3}$)} to get the list $L_{23}$. It  uses this list to get the intervals in the set $I_{u,v,w}$. Next, the algorithm iterates through the temporal edges in $L_1$. For every $e \in L_1$, if it discovers an interval in $I_{u,v,w}$ which contains $t$, it increments the value of in-count[$e$] by 1. Note that if $L_{1} \gets E_{w,v},$ then the algorithm sets $L_{2} \gets E_{w,u}$ and $L_{3} \gets E_{v,w}$. All the remaining steps will be the same.

However, observe that in this case, the algorithm iterates through the temporal edges $e \in L_1$ for every in-neighbor $x \in N^{-}_{\pi}(u) \cap N(v)$. To achieve the $O(m \alpha)$ bound, we need to avoid this from happening at all.

To get around this issue, we adopt the following approach: for every static edge $\{u,v\}$,  create two instances of a data structure that can be used store intervals. The first one, which we  denote by $D_{u,v}$, is used to deal with the temporal edges of the form $e = (u,v,t)$. The second one, denoted by $D_{v,u}$, is used to deal with temporal edges $e = (v,u,t)$. When the algorithm gets the list $I_{u,v,w}$, it inserts the intervals which belong to this list into $D_{v,w}$. Similarly, it inserts the intervals of the list $I_{u,w,v}$ in $D_{w,v}$. It repeats this procedure for all $(u,v) \in G^{\rightarrow}_{\pi}$ and $w \in N^{+}_{\pi}(u)$.

Finally, the algorithm iterates over the temporal edges in $G$. The hope is that for every temporal edge $e=(x,y,t)$, the data structure $D_{x,y}$ efficiently computes and returns the number of distinct lists $I_{a,x,y}$ which consist of at least one interval containing $t$ in it, as $a$ varies across $ N^{-}_{\pi}(x)$(or $N^{-}_{\pi}(y)$ if $\pi(y) < \pi(x)$) . In the next section, we offer a brief overview of how segment trees can be used for this purpose. A detailed discussion can be found in the supplementary material. The overall algorithm to handle in-neighbours is presented in algorithm-\ref{algo:eefa-overall-inneighbours}.  
\begin{algorithm}[t]
\caption{: \textsc{InNeighbours}($G, G_{S}, G^{\rightarrow}_{\pi}, \delta,$ in-count[$e$])}
\label{algo:eefa-overall-inneighbours}
\begin{flushleft}
\textbf{Input}: $G, G_{S}, G^{\rightarrow}_{\pi}, \delta$  \\
\begin{algorithmic}[1]
\State For every $\{u,v\} \in G_{S}$, build segment trees $T_{u,v}$ and $T_{v,u}$

\For{$(u,v) \in E(G^{\rightarrow}_{\pi})$}
    \For{$w \in N^{+}(u)$}
        \If{$w$ is adjacent to $v$}
            \State Get the lists $L_{2} \gets E_{v,u}$, $L_{3} \gets E_{w,u}$, $L_{2}^{'} \gets E_{w,u}$ and $L_{3}^{'} \gets E_{v,u}.$

            \State $L_{23} \gets $\textsc{FirstExceedingEntryLS}($L_{2}, L_{3}$)
            \State $L_{23}^{'} \gets $\textsc{FirstExceedingEntryLS}($L_{2}^{'}, L_{3}^{'}$)
            \State Get the list $I_{u,v,w}$ using $L_{23}$ and $L_{2}$
            \State Get the list $I_{u,w,v}$ using $L_{23}^{'}$ and $L_{2}^{'}$
            \State \textsc{InsertSegTree}($I_{u,v,w}, T_{v,w},u$)
            \State \textsc{InsertSegTree}($I_{u,w,v}, T_{w,v},u$)
        \EndIf
    \EndFor
\EndFor

\For{$\{u,v \} \in E(G_{S})$}
    \For{$e \in E_{u,v}$}
        \State count $\gets$ \textsc{LookUpSegTree}($t(e), T_{u,v}$)
        \State in-count[$e$] $\gets$ in-count[$e$] + count
    \EndFor
    \For{$e \in E_{v,u}$}
        \State count $\gets$ \textsc{LookUpSegTree}($t(e), T_{v,u}$)
        \State in-count[$e$] $\gets$ in-count[$e$] + count
    \EndFor
\EndFor
\end{algorithmic}
\end{flushleft}
\end{algorithm}
\subsection{Using Segment Trees to handle In-neighbors}\label{subsec:seg-trees-brief-discussion}
Fix a temporal edge $e = (u,v,t)$. Our goal is to compute in-count[$e$]. Let $T$ be the segment tree associated with the edge $(u,v)$. Assume that the intervals which were inserted in this tree came from three lists:  $L_1, L_2$ and $L_3$. Each node of $T$, in addition to a segment, stores a counter which may or may not be updated every time an interval is inserted in $T$. Note that by inserting an interval, we mean that we simulate the process of inserting an interval. The tree itself does not store any intervals.

When queried with $t$,  the algorithm looks for the segments in $T$ that contain $t$. All these segments lie on a path in $T$. The algorithm traverses this path, adds the values of the counters stored in the nodes on this path, and returns the resulting sum. And the hope is that this sum is equal to the number of lists among $L_1, L_2, L_3$, which contain at least one interval with $t$ in it.

For the sake of simplicity, assume that each of the three lists contain at least one interval with $t$ in it. Further, consider the simple procedure, in which every time an interval is inserted at a node $x$ in $T$, the procedure increments its counter by 1.  Ideally, if exactly one interval in each list contains $t$, then there is nothing to worry about, and the sum of counters correctly evaluates to 3. However, dealing with the case in which there exist multiple intervals containing $t$ within the same list is not as straightforward. 

We briefly describe the issue here. Let $x$ and  $y$ be two nodes on the path of the segments containing $t$. Assume $y$ to be a descendant of $x$ on this path. Suppose $L_1$ contains intervals $I_x$ and $I_y$, which will be stored at $x$ and $y$ respectively. If $I_x$ is inserted before $I_y$, then we can avoid increasing the counter of $y$. This is because when looking for $y$, the algorithm will encounter $x$ on its path. The moment is finds that that $x$'s counter is non-zero, it will prune the search and move on to the next interval.

However, if $I_y$ is inserted before $I_x$, then we have a problem. This is because now, the path leading upto $x$ does not contain $y$. Therefore, the algorithm will increment its counter by 1, and when queried with $t$, it will return 2 as opposed to 1. To get past this issue, we introduce the idea of \emph{lazy updating} of counters.

Broadly speaking, the idea is to delay counter updates rather than performing them immediately upon each interval insertion. This is implemented as follows: when an input interval is inserted and a node in the tree is selected to store it, instead of updating the node’s counter right away, we mark the node as grey. For example, both $x$ and $y$ would be marked grey in this step, which we refer to as \emph{Phase-1}.

Next, we proceed to \emph{Phase 2}, where all the intervals are re-inserted in the original order. During this re-insertion, if an interval is to be placed at node $k$, and none of the nodes along the insertion path are grey, we keep $k$ grey. However, if there is any grey node along the path to $k$, we reset $k$’s color to white. In the earlier example, since $x$ lies on the path to $y$ and is grey, $y$ is turned back to white.

In the final phase i.e. \emph{Phase 3}, the intervals are inserted once more, and this time, we update the counters only at the nodes that remain grey. So, in the previous case, we will only update the counter of $x$ and not of $y$. As a result, the algorithm will return 1 as opposed to 2. Full implementation details are provided in the supplementary material.

\begin{figure}
    \centering
     \begin{subfigure}{0.48\linewidth}
       \includegraphics[width=\linewidth]{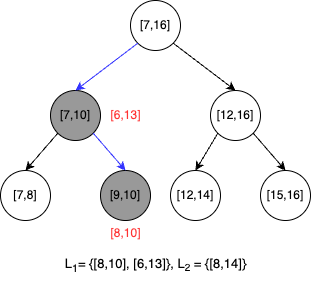}  
       \caption{Phase 1: Insert the intervals into the segment tree, update suitable colors but donot update the counters}
       \label{fig:seg-treep1}
    \end{subfigure}

    \begin{subfigure}{0.48\linewidth}
       \includegraphics[width=\linewidth]{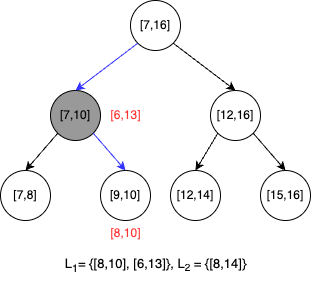}  
       \caption{Phase 2: Suppose we reinsert the intervals of $L_1$ in the same order. During the insertion of $[8,10]$, we come across the grey node containing the segment $[7,10]$. Hence we recolor the segment $[9,10]$, which is supposed to contain $[8,10]$ to white. On the contrary, since none of the nodes on the path leading up to segment $[7,10]$ were grey, we retain its color during the insertion of $[6,13]$. }
       \label{fig:seg-treep2}
    \end{subfigure}

    \begin{subfigure}{0.48\linewidth}
       \includegraphics[width=\linewidth]{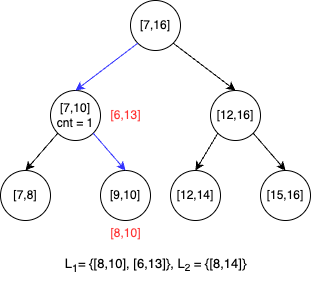}  
       \caption{Update the counters of grey nodes only.}
       \label{fig:seg-treep3}
    \end{subfigure}
   \caption{Insertion into Segment Trees, different phases}
    \label{fig:seg-trees-phases}
\end{figure}


\subsection{Putting it All Together}
Combining the ways to deal with the in-neighbours and out-neighbours, we present the overall algorithm in algorithm-\ref{algo:eefa-overall}.
Refer to Appendix for the proof.
\begin{theorem}\label{thm:overall-run-time}
    The overall running time of the algorithm is \\ $O(m \alpha \log \sigma_{\max})$
\end{theorem}
\section{$\exists \forall \forall $ Query and $\exists \forall \exists$ Query}
Given a temporal graph $G = (V, E)$, we seek to find solutions to the $\exists, \forall, \forall$ query:




Suppose $u$ solves the query. Then, for $\tau_{1} \%$ of $u$'s neighbors $v$, $G$ contains   at least one temporal edge $e = (u,v,t)$, which solves the "$\exists \exists \forall $" query, when its threshold is set to $\tau_{2}$. This observation leads to the following, very simple algorithm:
\begin{enumerate}
    \item First, solve for the $\exists \exists \forall_{\tau_{2}}$ query . Let $L$ denote the set of solutions to this query.
    \item For every vertex  $v \in V(G)$, create a dictionary $L_{v}$, indexed by the vertices in $N(v)$.
    \item For every $e = (u,v,t) \in L$, if $L_{u}[v] = 1$, then  proceed to the next edge in $L$. Otherwise, set $L_{u}[v] = 1$.
    \item Finally, compute $\sum_{v \in N(u)}L_{u}[v]$. 
    \item Finally, for every $v \in N(u)$, if $\sum_{v \in N(u)}L_{u}[v] \geq  \tau_{1}|N(u)|$, then include $u$ in the solution set.
\end{enumerate}
The algorithm for the $\exists \forall \exists$ is essentially the same, except for the lines (1) and (5). In particular, in line (1), instead of enumerating solutions to the $\exists \exists \forall$ query for a threshold value equal to $\tau_{2}$, the algorithm  enumerates edges $e$ which satisfy in-count[$e$] + out-count[$e$] $\geq 1$. The additional change needed is to update line (5)
 of the algorithm by replacing $\tau_{1}$ with $\tau$, since there is only one threshold.

\begin{algorithm}[t]
\caption{: \textsc{ExistExistForAll}($G, G_{S}, G^{\rightarrow}_{\pi}, \delta, \tau)$}
\label{algo:eefa-overall}
\begin{flushleft}
\textbf{Input}: $G, G_{S}, G^{\rightarrow}_{\pi}, \delta$  \\
\begin{algorithmic}[1]
\State Initialise in-count[] and out-count[] arrays of size $m$
\State $\text{Sol} \gets \phi$
\State \textsc{OutNeighbours}($G, G_{S}, G^{\rightarrow}_{\pi}, \delta$, out-count[$e$])
\State \textsc{InNeighbours}($G, G_{S}, G^{\rightarrow}_{\pi}, \delta$, in-count[$e$])
\For{$e = (u,v,t) \in E(G)$}
\If{out-count[$e$] + in-count[$e$]  $ \geq \tau |N(v)|$}
\State $\text{Sol} \gets \text{Sol} \bigcup \{e\}$
\EndIf
\EndFor
\State \Return $\text{Sol}$, $|\text{Sol}|$
\end{algorithmic}
\end{flushleft}
\end{algorithm}

\section{Experiments}

\begin{table}[t]
\centering
\scriptsize
\caption{Time to obtain solutions for $\exists \forall \exists$ query, across different choices of threshold, when $\delta =$ 4W}
\label{tab2:eae-runtime}
\begin{tabular}{|c|c|c|c|}
\hline
\textbf{Dataset} & \textbf{Threshold} & \textbf{Num. Solutions} & \textbf{Time(sec)} \\ \hline

\multirow{3}{*}{CollegeMsg (CM)} & 25\% & 178 & 0.10  \\ \cline{2-3}
& 50\% & 11 & 0.10  \\ \cline{2-3}
& 75\% & 0 &  0.10 \\ \cline{2-3}
\hline
\multirow{3}{*}{email-Eu-core-Temporal (ET)} & 25\% & 571 & 1.3  \\ \cline{2-3}
& 50\% & 338 &  1.2 \\ \cline{2-3}
& 75\% & 51 &  1.2 \\ \cline{2-3}
\hline
\multirow{3}{*}{sx-mathoverflow (MO)} & 25\% & 134 & 5.8  \\ \cline{2-3}
& 50\% & 20 &  5.9 \\ \cline{2-3}
& 75\% & 0 &  5.8\\ \cline{2-3}
\hline
\multirow{3}{*}{sx-superuser (SU)} & 25\% & 364 & 11.3  \\ \cline{2-3}
& 50\% & 95 &  11.1\\ \cline{2-3}
& 75\% & 4 &  11.3\\ \cline{2-3}
\hline
\multirow{3}{*}{sx-askubuntu (AU)} & 25\% & 408 & 5.2  \\ \cline{2-3}
& 50\% & 124 & 5.2 \\ \cline{2-3}
& 75\% & 5 &  5.3\\ \cline{2-3}
\hline
\multirow{3}{*}{wiki-talk-temporal (WT)} & 25\% & 4105 & 92.4  \\ \cline{2-3}
&  50\% & 1128 & 94.3 \\ \cline{2-3}
&  75\% & 108 &  92.4\\ \cline{2-3}
\hline
\multirow{3}{*}{sx-stackoverflow (SO)} & 25\% & 7279 & 1746.6  \\ \cline{2-3}
& 50\% & 1927 &1960.1   \\ \cline{2-3}
& 75\% & 123 &1684.7   \\ \cline{2-3}
\hline
\end{tabular}
\end{table}

\begin{table}[t]
\centering
\scriptsize
\caption{Time to obtain solutions for $\exists \exists \forall$ query, across different choices of threshold, when $\delta =$ 4W}
\label{tab1:eea-runtime}
\begin{tabular}{|c|c|c|c|c|c|}
\hline
\textbf{Dataset} & \textbf{Num. vertices} & \textbf{Num. edges} & \textbf{Threshold} & \textbf{Num Solutions} & \textbf{Time(sec)} \\ \hline

\multirow{3}{*}{CM} & \multirow{3}{*}{1.90E+03} & \multirow{3}{*}{5.98E+04} & 25\% & 102 & 0.06 \\ \cline{4-5}
& & & 50\% & 8 & 0.06 \\ \cline{4-5}
& & & 75\% & 0 &  0.05\\ \cline{4-5}
\hline
\multirow{3}{*}{ET} & \multirow{3}{*}{1.01E+03} & \multirow{3}{*}{3.32E+05} & 25\% & 7167 &  1.19 \\ \cline{4-5}
& & & 50\% & 478 &  0.84 \\ \cline{4-5}
& & & 75\% & 28&  0.35 \\ \cline{4-5}
\hline
\multirow{3}{*}{MO} & \multirow{4}{*}{8.86E+04} & \multirow{3}{*}{5.07E+05} & 25\% & 339 & 4.3  \\ \cline{4-5}
& & & 50\% & 65 & 2.8  \\ \cline{4-5}
& & & 75\% & 0 & 2.4 \\ \cline{4-5}
\hline
\multirow{3}{*}{SU} & \multirow{4}{*}{5.67E+05} & \multirow{3}{*}{1.44E+06} & 25\% & 1414 & 7.5  \\ \cline{4-5}
& & & 50\% & 383 & 6.07 \\ \cline{4-5}
& & & 75\% & 1 & 5.92 \\ \cline{4-5}
\hline
\multirow{3}{*}{AU} & \multirow{4}{*}{5.15E+05} & \multirow{3}{*}{9.64E+05} & 25\% & 1357 & 3.3 \\ \cline{4-5}
& & & 50\% & 398 & 3.03 \\ \cline{4-5}
& & & 75\% & 0 & 2.86\\ \cline{4-5}
\hline
\multirow{3}{*}{WT} & \multirow{4}{*}{1.14E+06} & \multirow{3}{*}{7.83E+06} & 25\% & 7118 & 51.7  \\ \cline{4-5}
& & & 50\% & 1183 & 43.8 \\ \cline{4-5}
& & & 75\% & 11 & 42.5  \\ \cline{4-5}
\hline
\multirow{3}{*}{SO} & \multirow{4}{*}{2.60E+06} & \multirow{3}{*}{6.35E+07} & 25\% & 25455 & 956.452 \\ \cline{4-5}
& & & 50\% & 5904  & 883.687 \\ \cline{4-5}
& & & 75\% & 1  & 840.509\\ \cline{4-5}
\hline
\end{tabular}
\end{table}

\begin{table}[t]
\centering
\scriptsize
\caption{Run time for SOTA triangle enumeration algorithm, $\delta$ = 4W}
\label{tab: BT-Runtime}
\begin{tabular}{|c|c|c|c|}
\hline
\textbf{Dataset} & \textbf{Num. vertices} & \textbf{Num. edges} &  \textbf{Time(sec)} \\ \hline

CM & 1.90E+03 & 5.98E+04 & 0.8\\  
\hline
ET & 1.01E+03 & 3.32E+05 & 10.9\\
\hline
MO & 8.86E+04 & 5.07E+05 & 1.4\\
\hline
SU & 5.67E+05 & 1.44E+06 & 5.12\\
\hline
AU & 5.15E+05 & 9.64E+05 & 4.2\\
\hline
WT & 1.14E+06 & 7.83E+06 & 141.29 \\
\hline
SO & 2.60E+06 & 6.35E+07 & 1227.9\\
\hline
\end{tabular}
\end{table}

\begin{table*}[t]
\centering
\scriptsize
\caption{Time to obtain solutions for $\exists \forall_{\tau_1} \forall_{\tau_2}$ query, across different choices of threshold, when $\delta =$ 4W.}
\label{tab3:eaa-runtime}
\begin{tabular}{|c|c|c|c|c|c|c|c|c|}
\hline
\textbf{Dataset} & \textbf{$\tau_{2}$} & \textbf{$\tau_{1}$} & \textbf{Num. Solutions} & \textbf{$\tau_{2}$} & \textbf{$\tau_{1}$} & \textbf{Num. Solutions} & \textbf{Time 1(sec.)} & \textbf{Time 2(sec.)} \\ \hline
\multirow{3}{*}{CollegeMsg} & \multirow{3}{*}{25\%} & 25\% & 0 & \multirow{3}{*}{50\%} & 25\% & 0 & 0.08 & 0.07 \\ \cline{3-4}\cline{6-7}
& & 50\% & 0 & & 50\% & 0 & 0.08& 0.07\\ \cline{3-4}\cline{6-7}
& & 75\% & 0 & & 75\% & 0 & 0.08& 0.07\\ \cline{3-4}\cline{6-7}
\hline
\multirow{3}{*}{email-Eu-core-temporal} & \multirow{3}{*}{25\%} & 25\% & 1 & \multirow{3}{*}{50\%} & 25\% & 0 & 1.2 & 0.8 \\ \cline{3-4}\cline{6-7}
& & 50\% & 0 & & 50\% & 0 &1.2 & 0.8\\ \cline{3-4}\cline{6-7}
& & 75\% & 0 & & 75\% & 0 &1.2 & 0.8\\ \cline{3-4}\cline{6-7}
\hline
\multirow{3}{*}{sx-mathoverflow} & \multirow{3}{*}{25\%} & 25\% & 0 & \multirow{3}{*}{50\%} & 25\% & 0 & 4.5 & 3.1 \\ \cline{3-4}\cline{6-7}
& & 50\% & 0 & & 50\% & 0 &4.4 & 3.1\\ \cline{3-4}\cline{6-7}
& & 70\% & 0 & & 75\% & 0 &4.5 & 3.1\\ \cline{3-4}\cline{6-7}
\hline
\multirow{3}{*}{sx-superuser} & \multirow{3}{*}{25\%} & 25\% & 4 & \multirow{3}{*}{50\%} & 25\% & 1 & 8.6 & 7.5 \\ \cline{3-4}\cline{6-7}
& & 50\% & 2 & & 50\% & 1 & 8.8&7.6 \\ \cline{3-4}\cline{6-7}
& & 75\% & 0 & & 75\% & 0 & 8.6& 7.5\\ \cline{3-4}\cline{6-7}
\hline
\multirow{3}{*}{sx-askubuntu} & \multirow{3}{*}{25\%} & 25\% & 7 & \multirow{3}{*}{50\%} & 25\% & 2 & 4.1 & 3.9 \\ \cline{3-4}\cline{6-7}
& & 50\% & 3 & & 50\% & 1 &4.2 & 3.9\\ \cline{3-4}\cline{6-7}
& & 75\% & 0 & & 75\% & 0 &4.1 & 3.9\\ \cline{3-4}\cline{6-7}
\hline
\multirow{3}{*}{wiki-talk-temporal} & \multirow{3}{*}{20\%} & 25\% & 60 & \multirow{3}{*}{50\%} & 25\% & 12 & 62.3 & 52.8 \\ \cline{3-4}\cline{6-7}
& & 50\% & 9 & & 50\% & 4 & 64.2& 53.5\\ \cline{3-4}\cline{6-7}
& & 75\% & 0 & & 75\% & 0 &62.5 &  52.7\\ \cline{3-4}\cline{6-7}
\hline
\multirow{3}{*}{sx-stackoverflow} & \multirow{3}{*}{25\%} & 25\% & 3737 & \multirow{3}{*}{50\%} & 25\% & 3 & 1195.35 & 936.01 \\ \cline{3-4}\cline{6-7}
& & 50\% & 1 & & 50\% & 1 &1191.53 &931.2 \\ \cline{3-4}\cline{6-7}
& & 75\% & 0 & & 75\% & 0 &1151.43 & 948.3\\ \cline{3-4}\cline{6-7}
\hline

\end{tabular}
\end{table*}

We evaluate our algorithm on a number of different graphs from SNAP repository \cite{snapnets}, with edge count ranging from 50K, to nearly 70M. We implement it in C++ and executed it on Intel Xeon Platinum 8380 CPU. For our experiments, we use algorithm-\ref{algo:practical-eea} to enumerate solutions to these queries.

In tables-\ref{tab1:eea-runtime}, \ref{tab2:eae-runtime} and \ref{tab3:eaa-runtime} (in appendix), we report the overall running time of our algorithm for thresholded $\exists \exists \forall$; $\exists \forall \exists$ and $\exists \forall \forall$ queries respectively. The run times, measured in seconds, are reported across various threshold values. For our experiments, we use algorithm-\ref{algo:practical-eea} to determine solutions to these triadic queries. 

\begin{algorithm}[t]
\caption{: \textsc{PRACTICAL-EEA}($G, G_{S},  \delta,$ count$[]$, $\tau$)}
\label{algo:practical-eea}
\begin{flushleft}
\textbf{Input}: $G, \delta,$ count[$e$], $\tau$  \\
\begin{algorithmic}[1]
\For{$\{u,v \} \in E(G_{S})$}
    \State Determine the lower degree end of the edge $\{u,v\}$. Let us denote the lower degree endpoint of this edge by $x$ and the other end by $y$.
    \For{$w \in N(x)$}
        \If{$w \in N(y)$}
            \State Get the lists $L_{1} \gets E_{u,v}, L_{2} \gets E_{u,w}$ and $L_{3} \gets E_{v,w}$. Also, get the lists $L_{1}^{'} \gets E_{v,u}, L_{2}^{'} \gets E_{v,w}$ and $L_{3}^{'} \gets E_{u,w}$ 
            \State $L_{12} \gets $\textsc{FirstExceedingEntry}($L_{1}, L_{2}$)
            \State $L_{23} \gets $\textsc{FirstExceedingEntry}($L_{2}, L_{3}$)
            \State $L_{12}^{'} \gets $\textsc{FirstExceedingEntry}($L_{1}^{'}, L_{2}^{'}$)
            \State $L_{23}^{'} \gets $\textsc{FirstExceedingEntry}($L_{2}^{'}, L_{3}^{'}$)\\
            \For{$e \in L_{1}$}
                \If{$t(e) \leq t(L_{12}[e]) \leq t(L_{23}[L_{12}[e]]) \leq t(e) + \delta$}
                    \State count[$e$] $\gets$ count[$e$] + 1
                \EndIf
            \EndFor
            \For{$e \in L_{1}^{'}$}
                \If{$t(e) \leq t(L_{12}^{'}[e]) \leq t(L_{23}^{'}[L_{12}^{'}[e]]) \leq t(e) + \delta$}
                    \State count[$e$] $\gets$ count[$e$] + 1
                \EndIf
            \EndFor
        \EndIf
    \EndFor
\EndFor
\For{$e \in E(G)$}
    \If{count$e$ $\geq \tau |N(v)|$}
        \State counter += 1\\
    \EndIf
\EndFor
\Return counter
\end{algorithmic}
\end{flushleft}
\end{algorithm}



 Our code is available at \footnote{Code available at \url{https://anonymous.4open.science/r/Triadic-FOL-Queries-CEF3}}.

For an $\exists \forall \forall$ query, we define two thresholds. We use $\tau_{1}$ and $\tau_{2}$ to denote the thresholds on the first and the second $\forall$ quantifiers respectively. 
We report the time taken to execute these queries in table-\ref{tab3:eaa-runtime}(in supplementary material due to lack of space). It is evident from tables \ref{tab2:eae-runtime}, \ref{tab1:eea-runtime}, and \ref{tab3:eaa-runtime} that all three algorithms take nearly the same amount time to produce solutions to these queries. Further, we also compare our algorithm against the state of the art triangle enumeration algorithm (BT) \cite{MinTemporalMotif} (refer table-\ref{tab: BT-Runtime}), and observe that our algorithm for the $\exists\exists\forall$ and $\exists \forall \forall$ queries is consistently faster than BT.  Since rest of the queries build upon the data structures output by the algorithm for $\exists \exists \forall$ query, they take a bit longer, but are reasonably fast i.e have comparable run time as BT. We note that the BT runtimes reported below account for the time it takes to enumerate the triangles. To obtain solutions to the above queries, one needs some indexing mechanism on top of the triangle enumeration subroutine. Therefore, BT will take at least the reported time to determine solutions to these queries.
\subsection{Insights} 
To gain some insights into the trend  of number of solutions w.r.t different parameter, we define our query as follows: $\exists u \exists v \in N(u) \forall_{\tau}N(u) \cap N(v) \phi(u,v,w)$, where $\phi$ represents the temporal triangle in Figure-\ref{fig:actual-temp-triangle}. We make similar modifications for $\exists \forall_{\tau_1} \forall_{\tau_2}$ and $\exists \forall_{\tau} \exists$ queries. The primary reason for this modification is to get enough solutions that will allow us to gain some insights. This also demonstrates that are techniques are simple, and versatile enough to accommodate various other sets. Further, we see that conclusions drawn from these plots are often consistent with the case in which the universe is just $N(v)$.  Note that there is no difference in the time it takes to compute the solutions to the queries when we change the universe from $N(v)$ to $N(u) \cap N(v)$. Doing so involves changing  line-18 of algorithm-\ref{algo:practical-eea} to $\tau |N(v) \cap N(u)|$. Further, in certain applications, one may be interested in predicated where the universe is indeed $N(v) \cap N(u)$. For instance, when trying to fetch cohesive groups of vertices, it may be more meaningful to have the intersection in the predicate.
\begin{figure}
    \centering
    \begin{subfigure}{0.48\linewidth}
       \includegraphics[width=\linewidth]{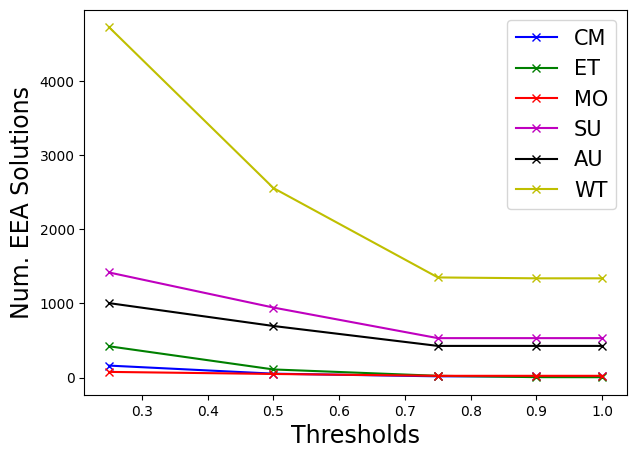}  
       \caption{$\exists \exists \forall$, $\delta$=1Hr.}
       \label{fig:eea-delta}
    \end{subfigure}
    \begin{subfigure}{0.48\linewidth}
       \includegraphics[width=\linewidth]{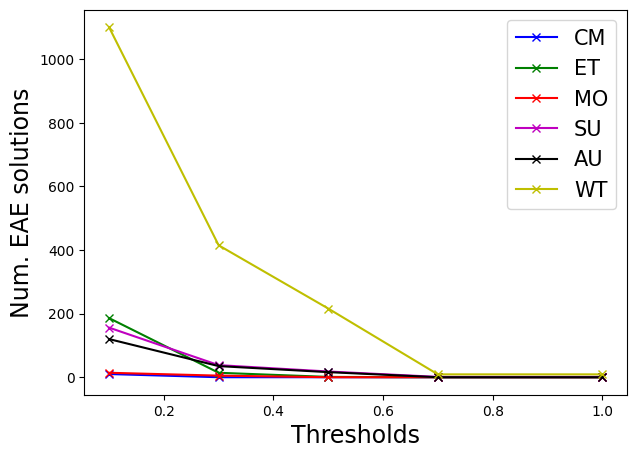}  
       \caption{$\exists \forall \exists$, $\delta$=1Hr.}
       \label{fig:eae-delta}
    \end{subfigure}
    \begin{subfigure}{0.48\linewidth}
       \includegraphics[width=\linewidth]{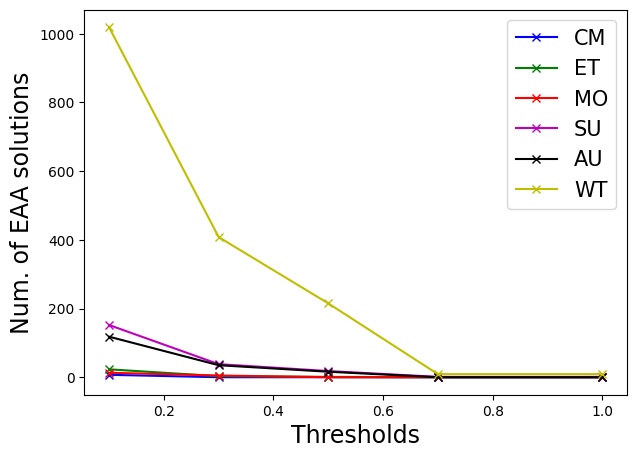}  
       \caption{$\exists \forall \forall$, $\delta$=1Hr, $\tau_{2} = 20\%$.}
       \label{fig:eaa-delta-thresh2-20}
    \end{subfigure}
    \begin{subfigure}{0.48\linewidth}
       \includegraphics[width=\linewidth]{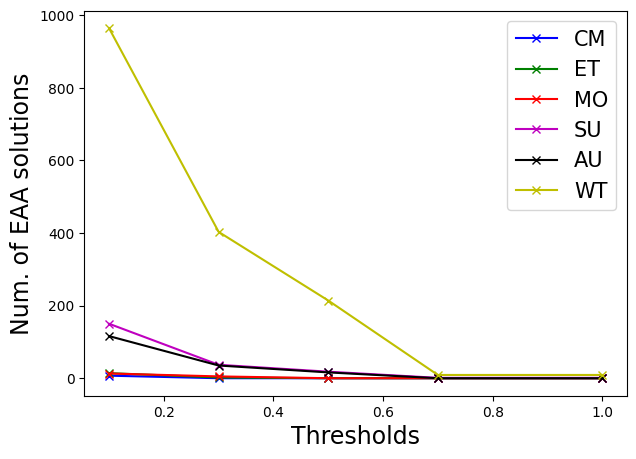}  
       \caption{$\exists \forall \forall$, $\delta$=1Hr, $\tau_{2} = 25\%$.}
       \label{fig:eaa-delta-thresh2-25}
    \end{subfigure}
    \caption{Number of solutions to different queries for a fixed delta, and varying thresholds}
    \label{fig:vary-thresh-fix-delta}
\end{figure}

\begin{figure}
    \centering
    \begin{subfigure}{0.48\linewidth}
       \includegraphics[width=\linewidth]{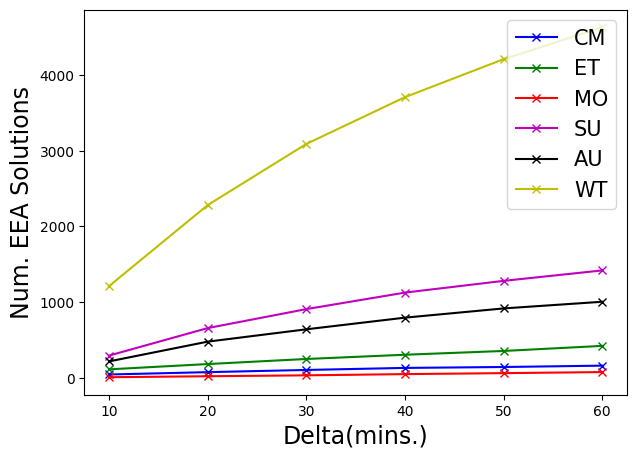}  
       \caption{$\exists \exists \forall$, $\tau$=25\%.}
       \label{fig:eea-thresh25}
    \end{subfigure}
    \begin{subfigure}{0.48\linewidth}
       \includegraphics[width=\linewidth]{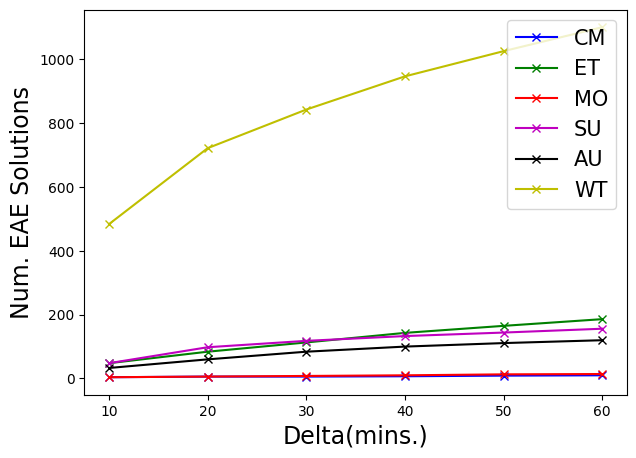}  
       \caption{$\exists \forall \exists$, $\tau$=10\%.}
       \label{fig:eae-thresh10}
    \end{subfigure}

    \begin{subfigure}{0.48\linewidth}
       \includegraphics[width=\linewidth]{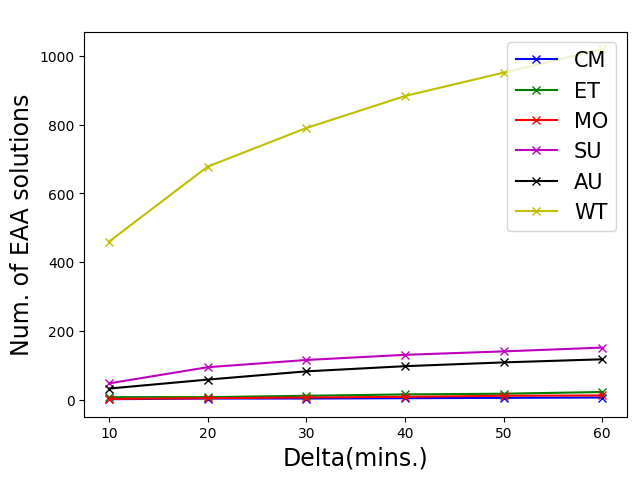}  
       \caption{$\exists \forall \forall$, $\tau_{1}$=10\%, $\tau_{2} = 20\%$.}
       \label{fig:eaa-thresh1-10-thresh2-20}
    \end{subfigure}
    \begin{subfigure}{0.48\linewidth}
       \includegraphics[width=\linewidth]{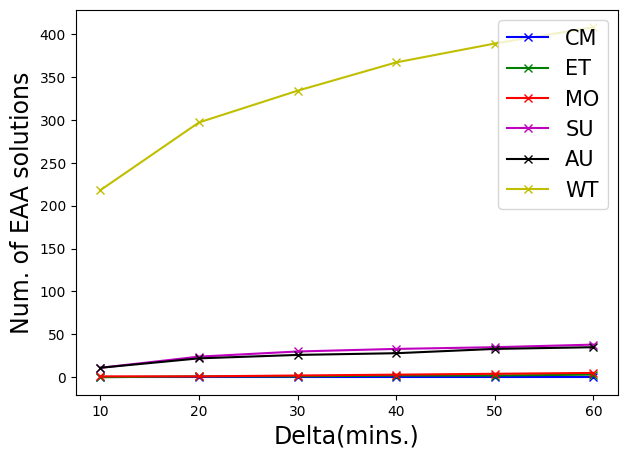}  
       \caption{$\exists \forall \forall$, $\tau_{1}$=30\%, $\tau_{2} = 20\%$.}
       \label{fig:eaa-thresh1-30-thresh2-20}
    \end{subfigure}

    \caption{Number of solutions to different queries for a varying delta, and fixed thresholds}
    \label{fig:vary-delta-fix-thresh}
\end{figure}

    \noindent \textbf{Number of solutions for a fixed $\delta$, varying threshold:} Figure-\ref{fig:eea-delta} shows the trend in the number of solutions to $\exists \exists \forall_{\tau}$ query as we vary the value of threshold. We fix $\delta$ to 1Hr. Observe that generally, as the threshold increases, the number of solutions tends to decrease. This is because any edge  that satisfies the $\exists \exists \forall_{\tau_{1}}$ , will also satisfy  $\exists \exists \forall_{\tau_{2}}$ for any $\tau_{2} < \tau_{1}$. For instance,  For instance, if a temporal edge $e = (u,v,t)$ forms a $\delta$-temporal triangle with at least 50\% of $u$ and $v$'s neighbors, it naturally satisfies the condition for 25\% as well. Therefore, the number of solutions for threshold $\tau_2$ is always at least as large as for any $\tau_1 \geq \tau_2$, explaining the observed decline in solutions with increasing thresholds.

    Further, observe  from Figure- \ref{fig:eea-delta}   that for relatively higher threshold values (typically exceeding 75$\%$), the number of solutions is almost constant. This implies that most  solutions to $\exists \exists  \forall_{\text{0.75}}$ query will also satisfy $\exists \exists \forall_{\text{0.9}}$ query.

    Similar trends can be observed in Figure-\ref{fig:eae-delta}, which depicts the variation in the number of solutions to $\exists \forall_{\tau} \exists$ query across a multitude of thresholds, when $\delta$ is set to 1Hr. Similar reasoning explains the decrease in the number of solutions  as the value of the threshold  increases. Further, beyond the 70\% threshold, the number of solutions is almost constant. In fact, it is 0 in most cases, except for the "wiki-talk-temporal" dataset, where it is 9. 

    Figures-\ref{fig:eaa-delta-thresh2-20} and \ref{fig:eaa-delta-thresh2-25} illustrate the trend in the number of solutions to the $\exists \forall_{\tau_{1}} \forall_{\tau_{2}}$ query for different values of the thresholds. As before, we set $\delta$ to 1Hr. In Figure-\ref{fig:eaa-delta-thresh2-20} we fix $\tau_{2}$ to 20\%, and in Figure-\ref{fig:eaa-delta-thresh2-25},  we set $\tau_{2}$ to 25\%.  Note that for a fixed $\tau_{1}$, the number of solutions to the $\exists \forall_{\tau_{1}} \forall_{0.2}$ query is at least the the number of solutions to the $\exists \forall_{\tau_{1}} \forall_{0.25}$ query. 

    Also note that generally, the number of solutions to $\exists \forall_{\tau} \exists$ query  exceeds the number of solutions to $\exists \forall_{\tau} \forall_{\tau^{'}}$ query. This is because every solution to $\exists \forall_{\tau} \forall_{\tau^{'}}$ query  naturally satisfies $\exists \forall_{\tau} \exists$ query.

    We note that similar observations can be extrapolated for the situations in which the universe is $N(v)$ rather than the intersection of neighborhoods.
    \begin{figure}
    \centering
    \begin{subfigure}{0.48\linewidth}
       \includegraphics[width=\linewidth]{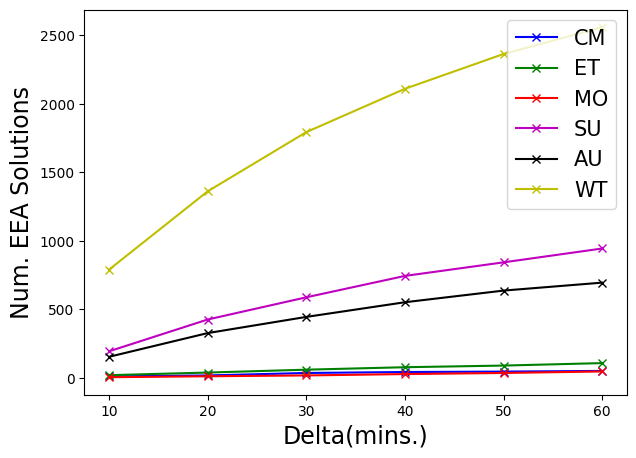}  
       \caption{$\exists \exists \forall$, $\tau$=50\%.}
       \label{fig:eea-thresh50}
    \end{subfigure}
    \begin{subfigure}{0.48\linewidth}
       \includegraphics[width=\linewidth]{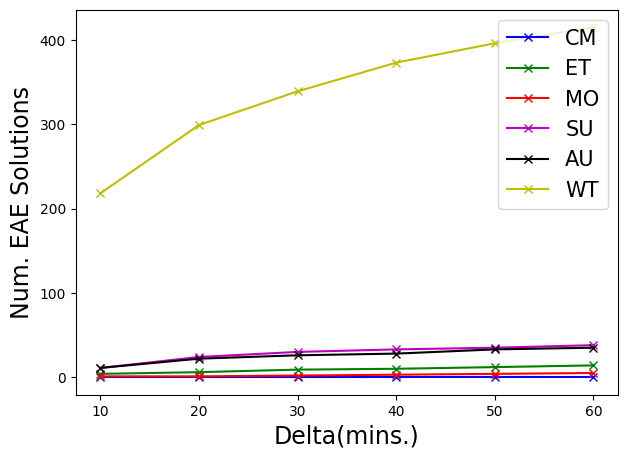}  
       \caption{$\exists \forall \exists$, $\tau$=30\%.}
       \label{fig:eae-thresh30}
    \end{subfigure}

    \begin{subfigure}{0.48\linewidth}
       \includegraphics[width=\linewidth]{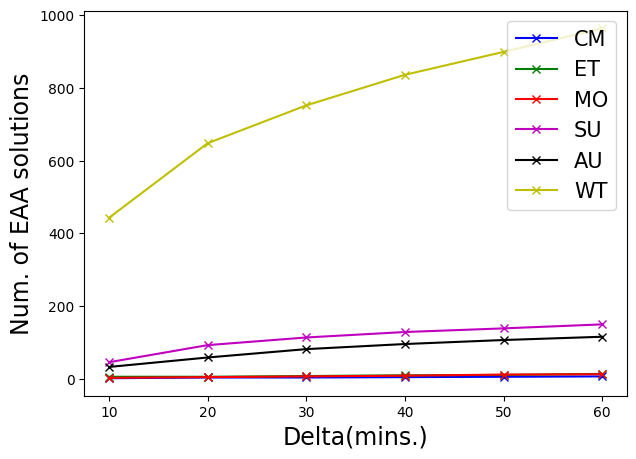}  
       \caption{$\exists \forall \forall$, $\tau_{1}$=10\%, $\tau_{2} = 25\%$.}
       \label{fig:eaa-thresh1-10-thresh2-25}
    \end{subfigure}
    \begin{subfigure}{0.48\linewidth}
       \includegraphics[width=\linewidth]{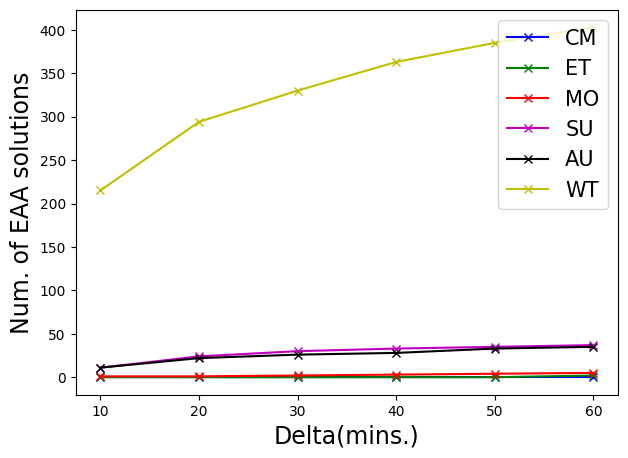}  
       \caption{$\exists \forall \forall$, $\tau_{1}$=30\%, $\tau_{2} = 25\%$.}
       \label{fig:eaa-thresh1-30-thresh2-25}
    \end{subfigure}
    
    \caption{Number of solutions to different queries for a varying delta, and fixed thresholds}
    \label{fig:vary-delta-fix-thresh-appendix}
\end{figure}

    \noindent \textbf{Number of solutions for a fixed threshold, varying delta:} Figures-\ref{fig:eea-thresh25} and \ref{fig:eea-thresh50} demonstrate the trend in the number of solutions to $\exists \exists \forall_{\tau}$ query as we vary the value of $\delta$. In Figure-\ref{fig:eea-thresh25}, we set the threshold to 25\% and in Figure-\ref{fig:eea-thresh50}, threshold is set to 50\%. Observe that the number of solutions to the query decreases as we increase $\delta$. 
    This is because, every temporal edge $e = (u,v,t)$ that satisfies the $\exists \exists \forall_{\tau}$ query for a given $\delta$ will naturally satisfy the $\exists \exists \forall_{\tau}$ for any $\delta^{'} \geq \delta$.  This explains the increase in the number of solutions with increasing $\delta$.

    Similar trends can be observed in the solutions to $\exists \forall_{\tau} \exists$ query (Refer figures-\ref{fig:eae-thresh10} and \ref{fig:eae-thresh30}). In figures \ref{fig:eae-thresh10} and \ref{fig:eae-thresh30}, we fix the thresholds to 10\% and 30\% respectively. As in the previous case, the number of solutions increases with increasing $\delta$ values.

    Figures-\ref{fig:eaa-thresh1-10-thresh2-20}, \ref{fig:eaa-thresh1-10-thresh2-25}, \ref{fig:eaa-thresh1-30-thresh2-20} and \ref{fig:eaa-thresh1-30-thresh2-25} demonstrate the variation in the number of solutions to the $\exists \forall \forall$ query as $\delta$ changes from 10mins. to 1Hr. In figures-\ref{fig:eaa-thresh1-10-thresh2-20} and \ref{fig:eaa-thresh1-30-thresh2-20}, we fix the value of $\tau_{2}$ to 20\% and plot the trend in the number of solutions for $\tau_{1} = 10\%$ and $\tau_{2} = 30\%$. In Figures-\ref{fig:eaa-thresh1-10-thresh2-25} and \ref{fig:eaa-thresh1-30-thresh2-25}, the value of $\tau_{2}$ is set to 25\% and we plot the number of solutions when $\tau_{1} = 10\%$ and $\tau_{2} = 30\%$. 
    
\noindent \textbf{Discovering structure using FOL queries:}
In the introduction, we saw how $\exists \exists \forall$ queries can be used to discover cohesive edges in a citation network. These queries automatically find edges that support a large number of triangles, and we discover that the corresponding papers are in the same topic.

\section{Conclusion}
In this work, we initiate the study of thresholded triadic first-order logic queries in temporal networks. We present a practical algorithm with provable guarantees to find solutions to these queries. Note that our algorithm can be extended for other temporal triangles as well. In future, we would like to investigate if the techniques presented in this work can be generalised to other motifs and first-order logic quantifiers. 
\bibliographystyle{ACM-Reference-Format}
\bibliography{refs}
\newpage
\appendix
\section{Appendix}
\subsection{Segment-Trees to deal with in-neighbours}\label{subsec:segment-trees-eefa}
\subsubsection{Primer on Segment Trees:}\label{subsubsec:seg-trees-first-attempt-eefa}
Before we proceed, we specify that when talking about a temporal edge $e = (x,y,t)$, we will assume $\pi(x) < \pi(y)$ for the ease of presentation. The other case can be handled analogously. 

Recall that for every static edge $\{u,v\}$, the algorithm creates two segment trees, namely $T_{u,v}$, and $T_{v,u}$. A segment tree is a balanced binary tree, used to store intervals. Every node of this tree contains a segment. We will use \textsc{Seg}($v$) to denote the segment stored at $v$. For every $v$ in $T$, \textsc{Seg}($v$) is obtained by combining the segments stored in the left and right children of $v$. In particular, if \textsc{Seg(LeftChild($v$))} = $[a,b]$ and \textsc{Seg(RightChild($v$))} = $[c,d]$, then \textsc{Seg}$(v)$ = $[a,d].$  For our setting,  in addition to a segment, every node contains a counter, which we  denote by \textsc{Counter}($v$) and a vertex, denoted by \textsc{Vertex}($v$). Given a node $v$ in $T$ and a set of intervals $I$, the \textit{canonical subset} of $v$ consists of intervals in $I$, which contain \textsc{Seg}($v$), but do not  contain \textsc{Seg}(\textsc{parent}($v$)). In a segment tree, if the canonical subset of a node contains an interval $I$, then  $I$ is  stored at that node. Although $I$ may belong to the canonical subset of several nodes in $T$, it is well known that at any given level, there can be at most 2 nodes where the $I$ might be stored \cite{book:Berg-CG-book}.
 This  fact is crucial in establishing the $O(\log n)$ complexity for inserting an interval in a segment tree. In particular, given an interval $I$, the time needed to look for the nodes, whose canonical subset contains $I$, is $O(h)$, where $h$ is the height of the tree.  

Coming back to the algorithm, note that for every $(u,v) \in G^{\rightarrow}_{\pi}$ and $w \in N^{+}_{\pi}(u)$, the algorithm first computes a list $I_{u,v,w}$ of intervals, and then inserts them in  the segment tree $T_{v,w}$. Finally, given a temporal edge $e = (u,v,t)$, the algorithm traverses along a path in $T_{u,v}$. Every node on this path consists of a segment that contains $t$. We shall refer to this path as the \emph{suitable path for $t$ in $T_{u,v}$}. The algorithm accumulates the  values of \textsc{Counter}($v$) for every node $v$ on this path and subsequently returns the sum. The hope is that this value  represents the number of vertices in $N^{-}_{\pi}(u) \cap N(v)$, which form at least one $\delta$-temporal triangle with $e$. This is precisely what in-count[$e$] is expected to store. 

\textbf{\emph{First Attempt :}}

Consider the intervals in $I_{w,u,v} = L$ for some $w \in N^{-}_{\pi}(u) \cap N(v).$ Suppose the algorithm does the following: for every $I \in L$, it looks for the node(s) in $T_{u,v}$ whose canonical subset contains $I$. For every such node $x$, it increments \textsc{Counter}($x$) by 1.

\begin{figure}
    \centering
    \begin{subfigure}{0.48\linewidth}
       \includegraphics[width=\linewidth]{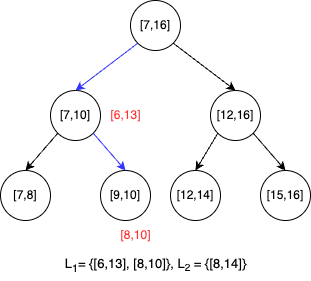}  
       \caption{Insertion of intervals into a segment tree}
       \label{fig:seg-tree1}
    \end{subfigure}

    \begin{subfigure}{0.48\linewidth}
       \includegraphics[width=\linewidth]{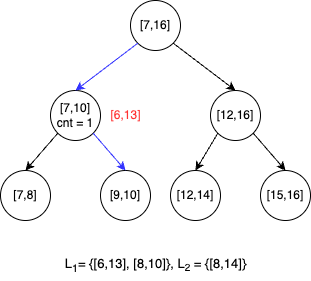}  
       \caption{Insertion into segment tree. The algorithm encounters a node with non-zero counter during the insertion of $[8,10]$, so it returns.}
       \label{fig:seg-tree2}
    \end{subfigure}

    \begin{subfigure}{0.48\linewidth}
       \includegraphics[width=\linewidth]{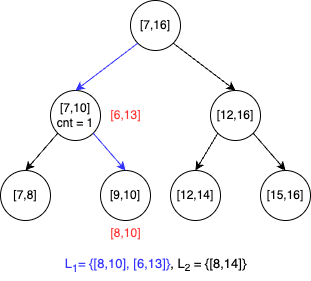}  
       \caption{If the orders of the intervals $[6,13]$ and $[8,10]$ are switched, we end up with the same issue.}
       \label{fig:seg-tree3}
    \end{subfigure}

    \caption{Insertion into Segment Trees}
    \label{fig:seg-trees}
\end{figure}

We will use Figure-\ref{fig:seg-tree1} as a running example. Suppose $I_{w,u,v} = L_{1} = \{[6,13], [8,10]\}.$ Observe that the intervals $[6,13]$ and $[8,10]$ belong to the canonical subset of nodes 1(i.e. node with segment $[7,10]$) and 2(i.e. node with segment $[9,10]$) respectively. As a result, after inserting the  intervals in $L_{1}$, the algorithm updates \textsc{Counter}($1$) and \textsc{Counter}(2) to 1. Now suppose $e = (u,v,9)$. 
 In order to compute in-count[$e$], the algorithm  looks for the suitable path of 9 in $T_{u,v}$. In this case, this path consists of nodes 1,2 and 4(i.e.  node with segment $[9,10]$). The algorithm adds the counters stored at these nodes and then returns the sum 2. However, both the intervals belong to the same list $L_{1}$. Therefore, ideally, the sum should have been 1 (to account for the in-neighbour $w$). More concretely, the algorithm has over-counted the contribution of $w$.

\textbf{\emph{Case when $I_{1}$ inserted before $I_{2}$:}}

Now suppose the algorithm does the following: for every node $v$ encountered during the insertion of an interval $I$, the algorithm checks if \textsc{Counter}($v$) is greater than 0.  If \textsc{Counter}($v$) is indeed greater than 0, then it simply returns. Otherwise, it proceeds with the usual insertion process. 

We emphasize that the intervals are not actually stored in the segment tree. When we say that the algorithm "inserts an interval $I$", we mean that it simulates the process of inserting $I$. In particular, it looks for the nodes whose canonical subset contains $I$, and then updates their fields.

How does this modification to the insertion procedure help? Consider the same example as before (refer figure-\ref{fig:seg-tree2}). First, the algorithm  inserts $I_{1} = [6,13]$, and updates \textsc{Counter}(1) to 1. Next, during the insertion of $I_{2}=[8,10]$, the algorithm visits  node $1$. Since \textsc{Counter}(1) > 0, the algorithm simply returns and moves on to the next interval in the input list. Note that this procedure avoids the over-contribution of $w$ to in-count[$e$], where $e = (u,v,9)$. This is because there exists a unique node on the suitable path of 9, namely node 1, which reflects the contribution   due to $w$.

Why did it work? During the insertion of an interval $I$, Suppose the algorithm visits a node $x$ with \textsc{Counter}($x$) > 1 . We can conclude that 
\begin{enumerate}
    \item there exists an interval $J$ in the input list, which was inserted before $I$ and contains \textsc{Seg}($x$) and 
    \item $I$ belongs to the canonical subset of some node in the subtree rooted at $x$. 
\end{enumerate}

Now consider an edge $e = (u,v,t)$. Observe that for every node $y$ in the subtree rooted at $x$, \textsc{Seg}($y$) $\subseteq$ \textsc{Seg}($x$). Consequently, if $J$ contains \textsc{Seg}($x$), then $J$ also contains every segment in the subtree rooted at $x$. As a result, for any interval $I$ which belongs to the canonical subset of $y$, if $t \in I \cap \textsc{Seg}(y)$, then it follows that $t \in J$. Therefore, following the insertion of $J$, it is no longer necessary to insert the intervals, which belong to the canonical subset of the nodes in the subtree, rooted at $x$.

\textbf{\emph{What if $I_{2}$ is inserted before $I_{1}?$}}

Up until now, we assumed that the interval $I_{1} = [6,13]$ was inserted before $I_{2} = [8,10]$. However, now let us switch the order of the intervals in $L_{1}$ (refer figure-\ref{fig:seg-tree3}). In particular, suppose  $L_{1} = \{[8,10], [6,13] \}$. Now the algorithm begins with the insertion of $I_{1}$. Since  none of the counters associated with the nodes, visited by the algorithm, are greater than 1,  the algorithm  updates \textsc{Counter}(2) to 1. Next, it inserts $I_{2}$. Since node $1$ is an ancestor of node 2, and $I_{2}$ is in the canonical subset of node 1, it follows that none of the counters, stored at the nodes encountered during the insertion of $I_{2}$, are greater than 1. This causes \textsc{Counter}($1$) to increase by 1. This creates a problem, since the suitable path for $e = (u,v,9)$ will return 2 as the value for in-count[$e$]. As a consequence, the algorithm over-counts the contribution of $w$.

More concretely, the algorithm fails if an interval $I$, which belongs to the canonical subset of a node $x$, is inserted after an interval $J$, which belongs to the canonical subset of a node in the subtree, rooted at $x$. Ideally, we would like $I$ to be inserted before $J$, not the other way round.

To get around this issue, the insertion algorithm proceeds in phases. We introduce an additional \textsc{Color} field into the structure of every node in the tree. This field is initially white for all nodes.

\textbf{\emph{Phase 1:}} Let $L$ denote a list of input intervals and $T$ be a segment tree. For every  $I \in L$, the algorithm searches for the nodes in $T$,  whose canonical subset contains $I$. During this process, if it encounters a grey node, then it simply returns and proceeds to the next interval in $L$. On the other hand, if it successfully reaches a node $v$ whose canonical subset contains $I$, then it updates $v$'s color from white to grey. Note that it does not update any counters along the way. 

In the end, a node $x$ is colored grey if
\begin{enumerate}
    \item $L$ contains an interval $I$ which belongs to the canonical subset of $x$ and
    \item None of the nodes on the  unique path from the root of $T$ to $x$ were grey in color during the insertion of $I$. 
\end{enumerate}
In our example, at the end of phase 1, nodes 1 and 2 will be grey in color (refer figure-\ref{fig:seg-treep1}).

\textbf{\emph{Phase 2: }} We begin by briefly describing the intuition for this phase. As seen before, for every interval $I \in L$, if the algorithm updates the counters  associated with all  the nodes, whose canonical subset contains $I$, then it could return an over-estimate to in-count[$e$] for some $e \in E(G)$ (refer figure-\ref{fig:seg-tree3}). Therefore, the idea is to use the colors assigned to the nodes of the tree to help us decide, which counters need to be updated. The eventual goal is update the counters associated with the grey nodes only. However, we aren't quite there yet. For instance, consider figure-\ref{fig:seg-tree3} At the end of phase 1, if $I_{2}$ is inserted before $I_{1}$, then both 1 and 2 will be grey, which as we saw before, is not good.  

At the end of phase 1, few nodes of $T$ will be grey in color, while the rest will be white. Let $S$ be the subset of grey nodes in $T$, which are reachable from the root via an entirely white path. In this phase, the goal is to recolor all the grey nodes back to white, except for the nodes in $S$. Observe that for any two nodes $x,y \in S$, $x$ cannot be in the subtree rooted at $y$, and vice-versa. 

Recall that an interval is stored at the node(s) whose canonical subset contain that interval. Now fix some $x \in S$. Since the root-to-$x$ path consists only of white nodes, none of the intervals in $L$ belong to the canonical subset of any ancestor of $x$. Therefore, every interval in $L$ will  be inserted at a node, which is either in $S$, or in the subtree, rooted at some $y \in S$. Further, if $I$ belongs to the canonical subset of $x$, and $J$ belongs to the canonical subset of some descendant $z$ of $x$, then for any $t \in J \cap \textsc{Seg}(z)$ it immediately follows that  $t \in I$. Therefore,  it suffices to keep the nodes in $S$ grey (same as the case when \textit{$I_{1}$ is inserted before $I_{2}$}) and update only their counters.     This prevents over-counting the contribution of a particular in-neighbour while computing the in-counts.  

Consider what happens during the insertion of an interval $I$.
\begin{enumerate}
    \item If the algorithm encounters a grey node, it takes note of this visit and proceeds with the insertion process. In the end, if it reaches a white node, it simply returns and moves on to the next interval in $L$. On the other hand, if the algorithm reaches a grey node, then it changes the color of this node, back to white.
    \item If the search for $I$ ends in a grey/white node and every other node visited by the algorithm was white in color, then it retains that node's color and proceeds to the next interval.
\end{enumerate}
In the end, a node $x$ will be \emph{white} in color if either
\begin{enumerate}
    \item none of the intervals in $L$ belong to the canonical subset of $x$ or
    \item  $x$ has an ancestor whose canonical subset contains an interval $I \in L$ 
\end{enumerate}
In our example, at the end of phase 2, only node 1 will be grey in color. In fact, consider the suitable path $P$ of $t$ in $T_{u,v}$ for some temporal edge $e = (u,v,t)$. Let $v_{1}, v_{2}$ and $v_{3}$ denote the grey nodes on this path, in the decreasing order of their height.  At the end of phase 2, only $v_{1}$ will be grey, since it is the only grey node on $P$, which is reachable from the root via an entirely white path.  Further, this will holds irrespective of the order on which the intervals are inserted. Finally, we have phase 3.

\textbf{\emph{Phase 3:}} 
Once the algorithm has identified the nodes in $S$, now all it needs to do is suitably update the counters of these nodes by reinserting the intervals in $L$. 
In this phase, the algorithm re-inserts the intervals of $L$. As before, during the insertion of an interval $I$,
\begin{enumerate}
    \item If the algorithm encounters a grey node, or it visits a white node $v$ with \textsc{Counter}($v$) > 1, then it simply returns and moves on to the next input interval.
    \item On the other hand, if the algorithm ends in a grey node, then it changes the color of that node back to white and updates its counter by adding 1 to it.
\end{enumerate}

\textbf{\emph{Why the \textsc{Vertex} field in the node structure?}}

Suppose there exists a list $L_{2} = I_{z,u,v} = \{ [8,14]\}$. Observe that $I_{3} = [8,14]$ belongs to the canonical subset of node 2. Therefore, during the insertion of $I_{3}$, the algorithm will visit node 1. Since \textsc{Counter}(1) > 0, it will simply return. Therefore, the suitable path for $e = (u,v,9)$ will return 1 as opposed to 2, since $z$ and $w$ are 2 different vertices in $N^{-}_{\pi}(u) \cap N(v)$, which have the property that their corresponding lists $I_{w,u,v}$ and $I_{z,u,v}$, contain at least one interval with 9 in it. This is where the vertex field comes to the rescue.

For a node $v$ in $T$, \textsc{Vertex}($v$) keeps track of the most recent vertex  which caused \textsc{Counter}($v$) to be updated.

In particular, after inserting the intervals of $I_{w,u,v}$, suppose the algorithm needs to insert intervals in $I_{z,u,v}$ for some $z \neq w$ in phase 3. Given an interval $I \in I_{z,u,v}$, 
\begin{enumerate}
    \item If the algorithm encounters a grey node, then simply return and move on to the next interval of $I_{z,u,v}$.
    \item If the algorithm visits a white node $x$ with \textsc{Counter}($x$) > 0, then check for \textsc{Vertex}($x$). If \textsc{Vertex}($x$) $\neq z$, then continue with the insertion process. On the contrary, if \textsc{Vertex}($x$) = $z$, then return.
    \item If the algorithm reaches(i.e. ends in) a grey node $x$, then simply update its color to white, increment it's associated counter by 1 and set \textsc{Vertex}($x$) to $z$.
\end{enumerate}
Note that phases 1 and 2 will not change. The above procedure is used in phase 3.

\begin{figure}
    \centering
     \begin{subfigure}{0.48\linewidth}
       \includegraphics[width=\linewidth]{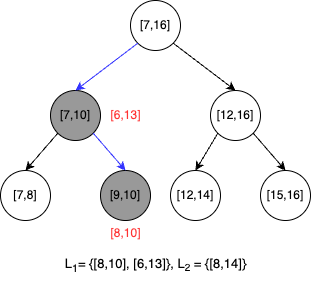}  
       \caption{Phase 1: Insert the intervals into the segment tree, update suitable colors but donot update the counters}
       \label{fig:seg-treep1}
    \end{subfigure}

    \begin{subfigure}{0.48\linewidth}
       \includegraphics[width=\linewidth]{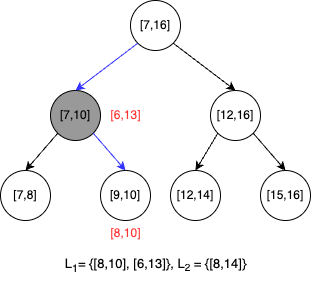}  
       \caption{Phase 2: Note the node with segment $[7,10]$ is the only grey node which is reachable from the root via an entirely white path. So we would like to retain its color. During insertion  of $[6,13]$, the algorithm encounters $[7,10]$, in which case it will update \textsc{FOUND} to true, go all the way to node containing $[8,10]$, and change its color back to white. }
       \label{fig:seg-treep2}
    \end{subfigure}

    \begin{subfigure}{0.48\linewidth}
       \includegraphics[width=\linewidth]{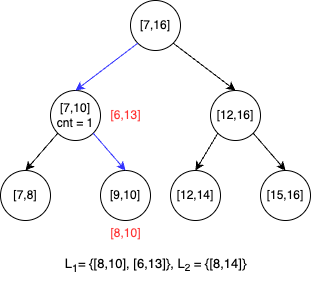}  
       \caption{Update the counters of grey nodes only.}
       \label{fig:seg-treep3}
    \end{subfigure}
   \caption{Insertion into Segment Trees, different phases}
    \label{fig:seg-trees-phases}
\end{figure}

Combining these ideas, the overall algorithm is presented in Algorithms-\ref{algo:eefa-overall}.

\textbf{\emph{How to build the Segment Tree?}}

All throughout, we assumed that each static edge comes with  pre-initialised segment trees. We now describe how to build the segment tree $T_{u,v}$. Note that in order to construct a segment tree, it is sufficient to specify the segments which will be stored in the leaves of the tree.

Typically, given a collection of intervals $L$, its segment tree is built using the distinct endpoints of the intervals in $L$ \cite{book:Berg-CG-book}. However, in our case, the number of such endpoints can be very large. Instead we build $T_{u,v}$(and $T_{v,u}$) using the timestamps of the temporal edges $(u,v,t)$(and $(v,u,t)$) for $t \in \mathbb{R}$. In particular, if $T = \{t_{1}, t_{2}, \dots, t_{r} \}$ denotes the collection of distinct timestamps of the temporal edges from $u$ to $v$, then the leaves of the segment tree will consist of the following segments $(t_{1}, t_{1}], (t_{1}, t_{2}], (t_{2}, t_{2}], (t_{2}, t_{3}], \dots, (t_{r}, t_{r}]$. 

Note that the intervals we wish to insert in $T_{u,v}$ may not necessarily have the same endpoints as timestamps in $T$. However given an interval $I$:
\begin{enumerate}
    \item If $I$ does not contain the timestamps in $T$, then $I$ will not cause any update in to the fields of the nodes, stored in $T_{u,v}$, which is fine.
    \item However, if $I$ does contain at least one of the timestamps in $T$, then using binary-search, we can find the largest  interval in $I$ with both endpoints in $T$. For eg: if $I$ contains $t_{i} \leq t_{i + 1} \leq \dots \leq t_{j}$, then using binary search, we can extract the interval $[t_{i}, t_{j}]$ and subsequently use it as a proxy for $I$ during the insertion process.   
\end{enumerate}

Putting it all together, the overall procedure is presented in Algorithm-\ref{algo:eefa-overall-inneighbours}.

\begin{theorem}\label{thm:in-neighbour-case-run-time}
The \textsc{InNeighbours}  subroutine runs in $O(m \alpha \log \sigma_{\max}$ time   
\end{theorem}
\begin{proof}
First, for every static edge $\{u,v\}$, the algorithm constructs a pair of segment trees. This takes $O(\sigma(u,v)\log \sigma_{\max})$ time. So the overall time to construct the segment trees is \\
$O(\sum_{\{u,v\} \in E(G_{S})}\sigma(u,v)\log \sigma_{\max}) = O(m \log \sigma_{\max}).$ 

  Now consider a edge $(u,v) \in G^{\rightarrow}_{\pi}$ and $w \in N^{+}_{\pi}(u)$. First, in line 5, the algorithm gets the lists $L_{1}, L_{2}$ and $L_{3}$. Then it invokes \textsc{FindExceedingEntryLS}, which takes $O(\sigma(u,v) + \sigma(u,w))$ time. Next, it inserts into the segment tree. Let $h_{\max}$ be the maximum height of the segment trees associated with the static edges in $G_{S}$. 

 Consider the subroutine \textsc{InsertSegTree}($I_{u,v,w}, T_{v,w}, u$). This subroutine inserts the intervals of the list $I_{u,v,w}$ in the segment tree $T_{v,w}$. Insertion of an interval in $T_{v,w}$ takes $O(h_{\max})$ time [refer chp. 10 of \cite{book:Berg-CG-book}]. Therefore, the overall time it takes to insert the intervals which belong to the list $I_{u,v,w}$ in $T_{v,w}$ is given by $O(|I_{u,v,w}|h_{\max})$. However, note that $|I_{u,v,w}| \leq \sigma(u,v)$.

 Putting these facts together, the time it takes for the algorithm to insert the intervals in the segment tree is \\ $O(\sum_{(u,v) \in E(G^{\rightarrow}_{\pi})}\sum_{w \in N^{+}_{\pi}(u)}\sigma(u,v)h_{\max})$. Note that $h_{\max} \leq \log \sigma_{\max}$. Therefore, the  overall run time is $O(m \alpha \log \sigma_{\max})$.

 Additionally, for every temporal edge $e$, getting the value of in-count[$e$] essentially involves looking up a path in a suitable segment tree. This can be done in time, proportional to the height of the tree. Therefore, the overall run time of lines 16-18 is $ O(m \log \sigma_{\max})$. 

 Combining all these together, the overall run time of the subroutine is $O(m \alpha \log \sigma_{\max}).$

 \end{proof}

\begin{algorithm}[t]
\caption{ \textsc{Phase1}($T, I$)}
\label{algo:eefa-insert-seg-tree-phase-1}
\begin{flushleft}
\textbf{Input}: $I$: interval, $T$: root of the segment tree $T$ \\
\begin{algorithmic}[1]
\If{\textsc{Color}($T$) = \textsc{Grey}}
   \State \Return
\EndIf
\If{\textsc{Seg}($T$) $\subseteq I$}
    \State \textsc{Color}($T$)$\gets$ \textsc{Grey}
    \State \Return
\EndIf
\If{$I \cap \textsc{Seg(LeftChild(}T)) \neq \phi$}
    \If{\textsc{Color(LeftChild(}$T)) = \textsc{Grey}$}
        \State \Return
    
    \Else
        \State \textsc{Phase1}(\textsc{LeftChild}($T$), $I$)
    \EndIf   
\EndIf
\If{$I \cap \textsc{Seg(RightChild(}T)) \neq \phi$}
    \If{\textsc{Color(RightChild(}$T)) = \textsc{Grey}$}
        \State \Return
    
    \Else
        \State \textsc{Phase1}(\textsc{RightChild}($T$), $I$)
    \EndIf   
\EndIf
\end{algorithmic}
\end{flushleft}
\end{algorithm}
\begin{algorithm}[t]
\caption{: \textsc{Phase2}($T, I$, \textsc{found})}
\label{algo:eefa-insert-seg-tree-phase-2}
\begin{flushleft}
\textbf{Input}: $I$: interval, $T$: root of the segment tree $T$, \textsc{found}: a boolean variable which is initially false \\
\begin{algorithmic}[1]
\If{\textsc{Seg}($T$)$\subseteq I$}
   \If{\textsc{found} = true}
        \State \textsc{Color}($T$) $\gets$ \textsc{White}
        \State \Return
    \EndIf
\EndIf

\If{$I \cap$ \textsc{Seg(LeftChild(}$T)) \neq \phi$}
    \If{(\textsc{Color(LeftChild(}$T))= $\textsc{Grey}) \textbf{or} (\textsc{Color(LeftChild(}$T))= $\textsc{White} \textbf{and} \textsc{Found} = true)}
        \State \textsc{Phase2(LeftChild(}$T),I,$true)
    \Else
        \State \textsc{Phase2(LeftChild(}$T),I,$false)
    \EndIf
\EndIf

\If{$I \cap$ \textsc{Seg(RightChild(}$T)) \neq \phi$}
    \If{(\textsc{Color(RightChild(}$T)) = $\textsc{Grey}) \textbf{or} (\textsc{Color(RightChild(}$T))= $\textsc{White} \textbf{and} \textsc{Found} = true)}
       \State \textsc{Phase2(RightChild(}$T),I,$true)
    \Else
        \State \textsc{Phase2(RightChild(}$T),I,$false)
    \EndIf
\EndIf
\end{algorithmic}
\end{flushleft}
\end{algorithm}

\begin{algorithm}[t]
\caption{: \textsc{Phase3}($T, I, w$)}
\label{algo:eefa-insert-seg-tree-phase-3}
\begin{flushleft}
\textbf{Input}: $I$: interval, $T$: root of the segment tree $T$, $w$: the in-neighbour being currently processed\\
\begin{algorithmic}[1]
\If{\textsc{Seg}($T$)$\subseteq I$}
   \If{(\textsc{Color}($T$) = \textsc{Grey})}
        \State \textsc{Color}($T$) $\gets$ \textsc{White}
        \State \textsc{Counter}($T$) $\gets$ \textsc{Counter}($T$) + 1
        \State \textsc{Vertex}($T$) $\gets w$
        \State \Return
    \EndIf
\EndIf

\If{$I \cap \textsc{Seg(LeftChild(}T)) \neq \phi$}
    \If{(\textsc{Color(LeftChild}($T$)) = \textsc{Grey}) \textbf{or} (\textsc{Counter(LeftChild}($T$)) > 0 \textbf{and} \textsc{Vertex(LeftChild}$(T))$ = $w$)} 
      \State  \Return
    \Else
    \State \textsc{Phase3(LeftChild(}$T), I, w)$
    \EndIf
\EndIf

\If{$I \cap \textsc{Seg(RightChild(}T)) \neq \phi$}
    \If{(\textsc{Color(RightChild}($T$)) = \textsc{Grey}) \textbf{or} (\textsc{Counter(RightChild}($T$)) > 0 \textbf{and} \textsc{Vertex(RightChild}$(T))$ = $w$)} 
      \State  \Return
    \Else
    \State \textsc{Phase3(RightChild(}$T), I, w)$
    \EndIf
\EndIf

\end{algorithmic}
\end{flushleft}
\end{algorithm}
\begin{algorithm}[t]
\caption{: \textsc{InsertSegTree}($L$,$T_{u,v}$,$w$)}
\label{algo:eefa-insert-seg-tree}
\begin{flushleft}
\textbf{Input}: $L = I_{w,u,v}$: list of input intervals, $T_{u,v}$: root of the segment tree associated with static edge $\{u,v\}$ and a vertex $w$\\
\begin{algorithmic}[1]
\For{$I \in L$}
    \State \textsc{Phase1}($T_{u,v},I$)
\EndFor
\For{$I \in L$}
    \State \textsc{Phase2}($T_{u,v},I$, false)
\EndFor
\For{$I \in L$}
    \State \textsc{Phase2}($T_{u,v},I,w$)
\EndFor
\end{algorithmic}
\end{flushleft}
\end{algorithm}

\begin{algorithm}[t]
\caption{: \textsc{LookUpSegTree}($t$,$T$)}
\label{algo:eefa-lookup-seg-tree}
\begin{flushleft}
\textbf{Input}: $t$: timestamp, $T$: root of a segment tree \\
\begin{algorithmic}[1]
\State sum $\gets$ 0
\If{$t \in $ \textsc{Seg}($T$)}
\State sum $\gets$ sum + \textsc{Counter}($T$)
\If{$t \in $\textsc{Seg(LeftChild($T$))}}
\State \textsc{LookUpSegTree}($t$,\textsc{LeftChild}($T$))
\Else
\State \textsc{LookUpSegTree}($t$,\textsc{RightChild}($T$))
\EndIf
\EndIf
\State \Return sum
\end{algorithmic}
\end{flushleft}
\end{algorithm}

\end{document}